\documentclass[sigconf, nonacm]{acmart}

\usepackage{cancel}

\newcommand{\pquote}[2]{\textit{``#1''}~{\small[#2]}}
\newcommand{\qquote}[1]{\textit{``#1''}}
\newcommand{\circled}[1]{\raisebox{.5pt}{(#1)}}

\definecolor{GREEN}{rgb}{0.31,0.65,0.18}
\definecolor{BLUE}{rgb}{0.06,0.62,0.84}
\definecolor{ORANGE}{rgb}{0.91,0.44,0.20}
\definecolor{CYAN}{rgb}{0.0,0.6,0.6}
\definecolor{PURPLE}{rgb}{0.6,0.0,0.6}
\newcommand{\Gen}{{\textcolor{BLUE}{\textbf{\textsc{Generate}}}}}
\newcommand{\Verif}{{\textcolor{GREEN}{\textbf{\textsc{Verify}}}}}
\newcommand{\Vary}{{\textcolor{ORANGE}{\textbf{\textsc{Vary}}}}}
\newcommand{\Iter}{{\textbf{\textsc{Iterate}}}}
\newcommand{\githublink}{\url{https://github.com/mario-michelessa/varifai}}

\usepackage[mathcal]{eucal}
\usepackage{bm}
\usepackage{multirow}
\usepackage{enumitem}
\usepackage{stfloats}

\AtBeginDocument{%
  \providecommand\BibTeX{{%
    \normalfont B\kern-0.5em{\scshape i\kern-0.25em b}\kern-0.8em\TeX}}}

\acmYear{2025} 
\setcopyright{cc}
 \setcctype{by}
 \acmConference[DIS '25]{Designing Interactive Systems Conference}{July 5--9,
 2025}{Funchal, Portugal}
\acmBooktitle{Designing Interactive Systems Conference (DIS '25), July 5--9,
 2025, Funchal, Portugal}\acmDOI{10.1145/3715336.3735847}
 \acmISBN{979-8-4007-1485-6/2025/07}

\begin{document}

\title[\texttt{Varif.ai} to Vary and Verify User-Driven Diversity in Scalable Image Generation]{\texttt{Varif.ai} to Vary and Verify User-Driven Diversity in Scalable Image Generation}

\author{Mario Michelessa}
\affiliation{
    \institution{National University of Singapore}
    \institution{Department of Computer Science}
    \city{Singapore}
    \country{Singapore}
}
\affiliation{
    \institution{IPAL}
    \city{Singapore}
    \country{Singapore}
}
\email{mario.michelessa@u.nus.edu}
\orcid{0009-0006-1489-6161}

\author{Jamie Ng}
\affiliation{
  \institution{Institute for Infocomm Research, A*STAR}
  \country{Singapore}
  \city{Singapore}
}
\email{jamie@i2r.a-star.edu.sg}
\orcid{0000-0003-2487-9251}

\author{Christophe Hurter}
\affiliation{
  \institution{Universite de Toulouse}
  \institution{ENAC}
  \city{Toulouse}
  \country{France}}
  \affiliation{
    \institution{IPAL}
    \city{Singapore}
    \country{Singapore}
}
\email{christophe.hurter@enac.fr}
\orcid{0000-0003-4318-6717}
\authornote{Co-corresponding author}

\author{Brian Y. Lim}
\affiliation{
  \institution{National University of Singapore}
  \institution{Department of Computer Science}
  \city{Singapore}
  \country{Singapore}
}
\email{brianlim@comp.nus.edu.sg}
\orcid{0000-0002-0543-2414}
\authornote{Co-corresponding author}

\begin{abstract}
Diversity in image generation is essential to ensure fair representations and support creativity in ideation. Hence, many text-to-image models have implemented diversification mechanisms. Yet, after a few iterations of generation, a lack of diversity becomes apparent, because each user has their own diversity goals (e.g., different colors, brands of cars), and there are diverse attributions to be specified. To support user-driven diversity control, we propose Varif.ai that employs text-to-image and Large Language Models to iteratively i) (re)generate a set of images, ii) verify if user-specified attributes have sufficient coverage, and iii) vary existing or new attributes. Through an elicitation study, we uncovered user needs for diversity in image generation. A pilot validation showed that Varif.ai made achieving diverse image sets easier. In a controlled evaluation with 20 participants, Varif.ai proved more effective than baseline methods across various scenarios. Thus, this supports user control of diversity in image generation for creative ideation and scalable image generation.
\end{abstract}



\begin{teaserfigure}
\centering
  \includegraphics[width=17.6cm]{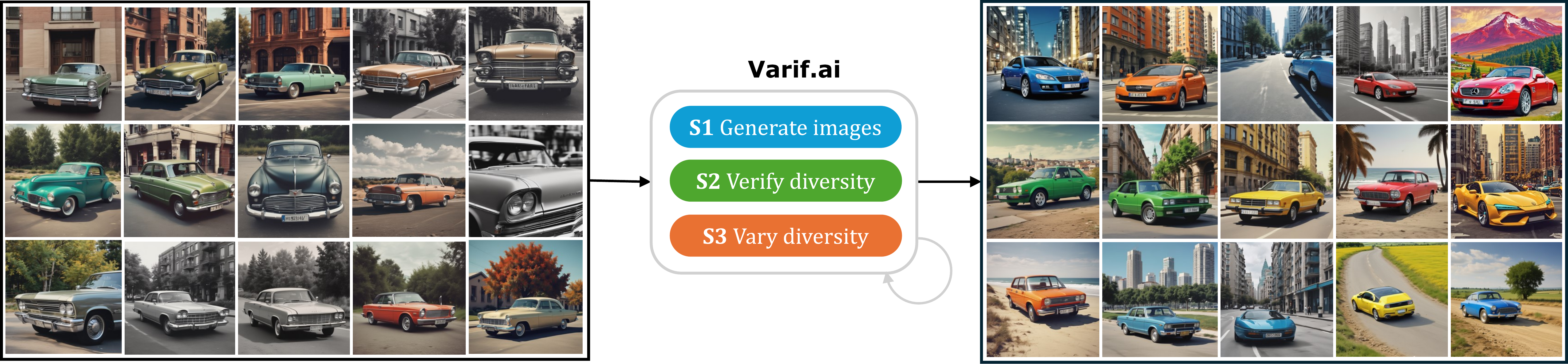}
  \caption{
  Varif.ai supports user-defined diversity to S1) generate images, S2) verify their diversity based on attributes, and S3) vary diversity iteratively.
  Left: Images generated with a state-of-the-art diffusion model~\cite{lin2024sdxl} using the prompt ``\textit{a picture of a car}'' lacks diversity. 
  Right: Images generated with Varif.ai has more diversity in car color, car model, and background landscape.
  }
  \Description{The figure showcases the Varif.ai system, which allows users to control diversity in generated image sets. The image is divided into three main sections. On the left, there is a grid of nine vintage-style car images, all with a similar color palette and aesthetic, suggesting limited diversity. In the center, the Varif.ai process is depicted with three steps: S1) Generate images, S2) Verify diversity, and S3) Vary diversity, connected in a flowchart. To the right, there is a second grid of nine car images, demonstrating more variation. These cars differ significantly in color, style, setting, and time period, illustrating increased diversity in the image set after using Varif.ai. The figure highlights the system's iterative process for improving image diversity.}
  \label{fig:teaser}
\end{teaserfigure}

\maketitle

\section{Introduction}

Generative text-to-image models have become highly popular for their ability to create high-quality images from natural language prompts~\cite{dhariwal2021diffusion, luccioni2024stable}, while offering users versatility and control over the generation process.
Various approaches have been developed to provide fine-grained control, including prompt engineering techniques~\cite{liu2022design, oppenlaender2023taxonomy, wang2023reprompt, brade2023promptify, feng2023promptmagician}, and direct manipulation techniques like in-painting~\cite{lugmayr2022repaint, wang2024promptcharm}, image editing~\cite{hertz2022prompt}, and layout control~\cite{zhao2024uni}.

However, current generative models often lack diversity.
For example, when generating a set of cars, generative models frequently exhibit biases toward specific car models, car colors, or landscapes (see Fig.\ref{fig:teaser}, left).
This can hinder creativity in design by limiting inspirational examples. 
The lack of diversity also extends beyond visual attributes and affects social ones, like gender~\cite{cho2023dall, wang2023t2iat}, geographic origin~\cite{basu2023inspecting, schramowski2023safe, zhang2024partiality}, or social background~\cite{bianchi2023easily, luccioni2024stable}.
Ensuring diversity in image collections is crucial not only for mitigating stereotypes and reducing biases~\cite{silva2023representation}, but also for supporting creative ideation for effective communication and design~\cite{herring2009getting, lee2010designing, dow2010parallel}.

Several augmentation methods have been proposed to automatically increase diversity~\cite{betker2023improving, hao2024optimizing}, but they cannot encompass all diversity, since this depends on user needs and perception. Hence, diversification needs to be user-driven.
With current techniques to control image generation, users can fine-tune each image, but this is tedious and not scalable to diversify a large set of images.

To support user-driven diversity,
we propose Varif.ai to empower users to define, measure, and improve the diversity of image sets through an iterative three-step process: 
S1) \textit{(Re)Generate} a set of images based on an initial prompt,
S2) \textit{Verify} whether the user-specified attributes have the desired diversity coverage,
S3) \textit{Vary} existing attributes and their labels, or introduce new attributes. 
Our approach is built on top of off-the-shelf foundation models, making it model-agnostic and not requiring model retraining.

We first conducted an elicitation study with 8 participants to determine the needs for diversity in image generation, and understand the limitations of standard prompt engineering.
From these initial findings, we developed and designed Varif.ai to help users to 
verify attribute diversity through histogram visualizations and 
vary diversity by interactively changing the histograms to execute probabilistic prompt generation.
Following this, we conducted a formative study with the same 8 participants to verify that Varif.ai improved their experience and ability to diversify images, and explore the diverse images they generated.
Next, we conducted a controlled evaluation study with 20 participants to compare Varif.ai against baseline methods (simple prompt engineering, and automated-diversified GPT-4o~\cite{achiam2023gpt} and Promptist~\cite{hao2024optimizing}), and examined how Varif.ai supports diversity with different specification degrees (open-ended, attribute-specific, attribute-label-specific).

Our \textbf{contributions} are:
\begin{enumerate}[label=\arabic*)]
    \item Insights into how users define, measure, and control diversity for generated images.

    \item Varif.ai, an interactive visualization\footnote{Code available at \githublink} 
    to generate diverse images through iterative verification and variation of attributes.
    
\end{enumerate}

Evaluations show Varif.ai improved user-driven image diversity.
We discuss its scope, limitations, generalization and future work.

\section{Related Work}
We discuss background work on text-to-image generative models and user interfaces, and methods to improve and evaluate diversity.

\subsection{Prompt Engineering for Image Generation}

Diffusion models~\cite{yang2023diffusion}, such as DALL-E 2~\cite{ramesh2022hierarchical}, Stable Diffusion~\cite{rombach2022high}, and SD-XL~\cite{podell2024sdxl}, have enabled the generation of remarkably high-quality images. 
Lay users can leverage these models through text-to-image prompt engineering.
However, prompting can be difficult and unintuitive~\cite{zamfirescu2023johnny}.
Liu and Chilton~\cite{liu2022design} emphasized the importance of iterating prompts multiple times, given the stochastic nature of the models. 
Oppenlaender~\cite{oppenlaender2023taxonomy} and Hutchinson et al.~\cite{hutchinson2022underspecification} highlighted the crucial role of detailed modifiers in prompts because images contain more information than simple text. 
While these approaches require human ingenuity to prompt effectively, an alternative approach 
is to leverage large language models (LLMs) to generate augmented prompts to improve image quality and fit~\cite{hao2024optimizing, betker2023improving}. 
In this work, we use a similar approach to augment diversity.

\subsection{Interaction Design for Image Generation}
Sidestepping the challenges in prompt engineering, researchers have proposed visualization-based interfaces for human-AI collaboration with text-to-image models. 
PromptCharm~\cite{wang2024promptcharm} integrates multimodal techniques, such as inpainting~\cite{lugmayr2022repaint} for modifying specific image regions, and attention-control mechanisms for tuning the intensity of concepts in images, surpassing traditional prompt engineering to better align with user preferences.
PromptPaint~\cite{chung2023promptpaint} allows users to go beyond language to mix prompts that express challenging concepts through paint-like interactions.
Once again, these tools focus on image quality rather than diversity.
Targeting artists~\cite{chung2023artinter}, designers~\cite{son2024genquery, liu2022opal, liu20233dall, peng2024designprompt}, and game developers~\cite{dang2023worldsmith}, some tools have been proposed to enhance creativity.
However, they focus on generating one image at a time, which can be tedious when diversifying a set of images together.

\subsection{Image Diversification}

Obtaining visual references is typically done via web search due to its accessibility and breadth. This has been widely adopted in 
image creativity tools, such as 
PicturePiper~\cite{fass2000picturepiper} with reconfigurable pipelines for interactive image retrieval, 
example-based design~\cite{lee2010designing} with adaptive galleries, and
MetaMap~\cite{kang2021metamap} for idea exploration  via multimodal semantic, color, and shape spaces. 
However, image search is limited to existing content, restricting the exploration of novel or hybrid concepts. 
Generative models offer a more expressive alternative~\cite{mozaffari2022ganspiration, son2024genquery}, allowing users to synthesize and steer images through fine-grained control over visual attributes. 

Yet, image generation introduces challenges such as distributional biases and limited support for user-driven diversity.
State-of-the-art text-to-image models have been shown to be biased in different attributes, such as gender~\cite{cho2023dall, wang2023t2iat}, geographic origin~\cite{basu2023inspecting, schramowski2023safe}, social background~\cite{bianchi2023easily, luccioni2024stable}, or cultural artifacts~\cite{zhang2024partiality}. 
Tests with lay-users also revealed that a lack of variability in AI-generated images significantly hinders creative outcomes~\cite{wadinambiarachchi2024effects}.
Several approaches have been proposed to improve diversity in generated images, based on  
model fine tuning~\cite{shen2023finetuning, miao2024training} or prompt optimization~\cite{zhang2023iti} to align with predefined target distributions for specific attributes, in the pursuit of fairness.
Ironically, some automatic approaches have sparked controversy of over-doing the diversity~\cite{geminiDiversity, publishedItSeemsAI2024, GoogleSaysIts2024}.
Indeed, user control is needed to manage the diversity to satisfy their specifications, which we support in our work.

\subsection{Measuring and Visualizing Diversity}
In order to improve diversity, one must first measure it.
Several computational metrics have been defined for diversity~\cite{asudeh2019assessing, xia2015learning, kaminskas2016diversity, petchey2002functional, lehman2011abandoning}, 
but as singular metrics, each does not encompass all definitions of diversity, and they are too abstract for lay users to understand.
In contrast, 
more qualitative approaches, such as examples~\cite{zhang2021method} or visualizations (e.g., histograms, clustering), offer richer, more accessible insights. 
Assogba et al.~\cite{assogba2023large} used clustering in a semantically meaningful embedding space to support exploration and multi-criteria evaluation of generated images.
IF-City~\cite{lyu2023if} employed bar charts and heatmaps on 3D urban plans to explain fairness and inequality.
FairSight~\cite{ahn2019fairsight} and FairVis~\cite{cabrera2019fairvis} used multiple coordinated views of various visualizations to allow inspection of data and distributional biases.
Inspired by these methods, we include computational metrics and visualizations to evaluate, measure, and communicate diversity in images as a key step in our work.

\section{Elicitation user Study}
\label{sec:elicitationStudy}

In an elicitation study, we determined the needs for diversity in image generation and the limitations of basic prompt engineering,

\subsection{Method}
\label{sec:study1method}
\subsubsection{Participants} We recruited 8 participants (E1–E8) through snowball sampling for the elicitation study. It was conducted over Zoom for 30 minutes. Participants were compensated \$22 USD in local currency. 
They were 5 females, median 28 years old, with varying experience in design and AI (see Table~\ref{tab:experts} for details). 

\begin{table*}[]
    \centering
    \caption{Background of participants indicating their experience in design and AI, and selected topic in the elicitation study.}
    \begin{tabular}{lp{17.0em}p{15.0em}p{12.5em}}
        \hline
          \textbf{Participant}&  \textbf{Design experience} & \textbf{Image generative AI experience} & \textbf{Self-chosen image topic} \\
        \hline
        \addlinespace[0.05cm]
         E1 (30M) & Professional artist for 4-5 years, mainly in paint. & None. & Poses of falling monkey. \\
        \addlinespace[0.1cm]
         E2 (24F) & Professional designer in advertising for 3 years, and children's book illustration. & None. & Poster of Asian friends traveling in Scotland. \\ 
        \addlinespace[0.1cm]
         E3 (23M) & Graduated in design, but no professional experience. & Some use of image generation AI. & Very tall red concrete building. \\ 
        \addlinespace[0.1cm]
         E4 (26F) & None. & None. & Outdoor community centers.\\ 
        \addlinespace[0.1cm]
         E5 (25F) & Professional designer and animator in film industry for 3 years. & None & Cockroaches drinking beer in a sewer. \\ 
        \addlinespace[0.1cm]
         E6 (24F) & Professional designer in textile industry for 4 years. & None & Dancing colorful frogs. \\ 
        \addlinespace[0.1cm]
         E7 (44F) & Professional Artist for 20 years. & Extensive use for ideation. & Ancient doorframe with intricate details. \\ 
        \addlinespace[0.1cm]
         E8 (27M) & None. & Some use of image generation AI & Logo for an AI startup \\ 
        \addlinespace[0.05cm]
         \hline
    \end{tabular}
    \label{tab:experts}
    \Description{Table summarizing the background of eight participants in a study, including their design experience, experience with image generative AI, and their self-chosen image topic. Design experience ranges from none to 20 years, with varying specialties. AI experience spans from none to extensive. Topics include scenes like a falling monkey, outdoor community centers, and a logo for an AI startup.}
\end{table*}

\subsubsection{Study procedure} 

The study consisted of two parts: 
i) semi-structured interview to understand participants' needs for diversity,
ii) user interaction with an image generation tool while thinking aloud to explore the limitations of prompt engineering.
The interview centered on the following tasks and questions:
\textit{Recall a time when you needed to browse for image references.
Explain how you did it and which tools you used.
How many images did you review before deciding on the right one?
How would you have liked to control the set of images proposed to you?}
Participants shared their screen with the researcher so they could show references if they wished.

Next we asked participants to use a basic tool that we developed with a text input to enter prompts and a gallery displaying 10 generated images. They were tasked to generate images for inspirational references for any topic (see topics in Table~\ref{tab:experts}).
They were asked to think aloud to articulate their rationale when writing the prompts and examining the generated images, and to understand how they considered diversity and measured and (wanted to) control it.

\subsection{Findings}
We performed the thematic analysis~\cite{terry2017thematic} on user interactions and utterances. 
Thematic codes were identified by the first authors and verified by the last author.
We found that participants
i) iteratively refined their goals during the ideation process, 
ii) considered diversity in terms of attributes and labels, and
iii) wanted to control diversity by defining labels and changing proportions.

\subsubsection{Iterative Ideation and Need for Diversity}
All participants emphasized the importance of having diverse images when selecting visuals. For example, in the context of creating communicative images for a presentation slide, E4 explained that he often needed to look at \qquote{20 images, sometimes more, to see a wide range of possibilities before choosing what to include}. He also noted the role of serendipity, adding that sometimes \qquote{I particularly like one of them and continue my search based on [that] image}.

Participants noted the iterative nature of their creative process as they refined or revisited their goals. 
E2 remarked that \qquote{the first [iteration] should be kind of unbiased, the first idea should be very rough and very draft[-like]}. 
Similarly, E7 shared that \qquote{at the start I want to have references for different ideas quickly [...] and little by little the exploration will be narrower and narrower, the more I add criteria}. 
E8 remarked that \qquote{it happens that my initial searches were [sometimes] better, so I revert back to [them]. [...] I always keep the first search close so that I can compare and revert to it}.

\begin{figure*}[t]
  \centering
  \includegraphics[width=\linewidth]{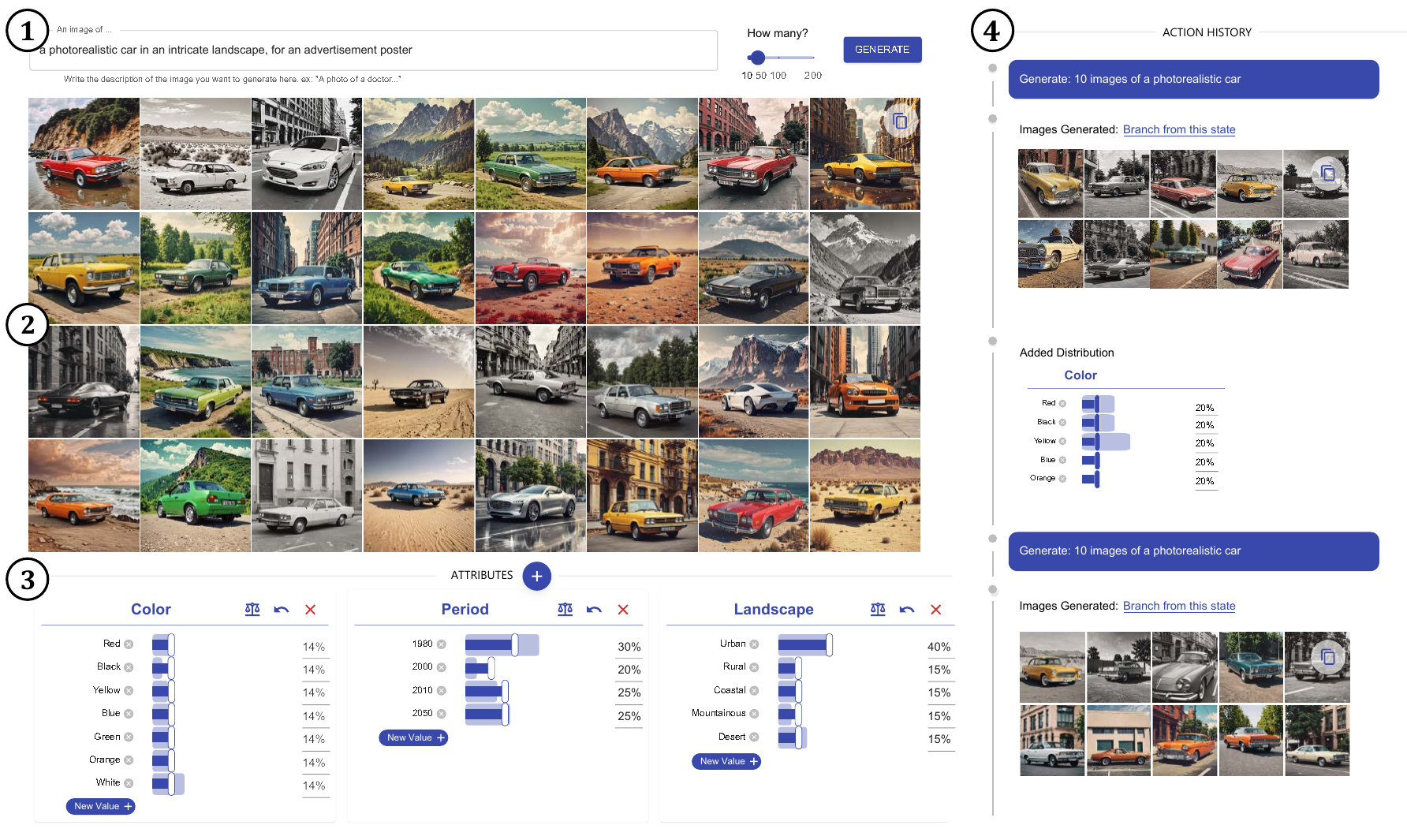}
  \caption{The user interface of Varif.ai comprises four main panels. The Prompt text input \circled{1} allows users to define a general description of the images and specify the number of images to generate. The Image gallery \circled{2} displays the generated images and provides popup tooltips showing their specific descriptions and attribute labels. The Attributes panel \circled{3} allows users to add attributes (e.g., Color, Period, Landscape), measure the distributions of their labels (e.g., Red, Black, etc., for Color) as vertical histograms, and modify the distributions as desired. The Action history panel \circled{4} records the generated images after each iteration as the user modifies the attributes. }
  \label{fig:interface}
  \Description{The interface consists of four key sections: Input Section: Users can enter a prompt (e.g., "a car") and choose the number of images to generate with a slider (10-200) and a "Generate" button. Image Grid: Displays a variety of generated car images, showcasing different colors, settings, and models. Attributes Panel: Allows users to adjust attributes like "Color," "Period," and "Landscape" through sliders to refine the diversity of the image set. Action History: Tracks user actions, displaying previous image sets and attribute adjustments (e.g., color and time period) with bar graphs. Users can also download earlier versions of the generated images.}
\end{figure*}

\subsubsection{Definition of Diversity}

Participants sought for diversity in terms of attributes, often specifying several labels. 
Five out of 8 participants mentioned the style of the images, providing various desired labels: E2 mentioned \qquote{abstract, steampunk, realistic}, E5 suggested \qquote{cartoony, dark, cute}, and E1 referred to \qquote{pixel-art, video game, photography}.
Other frequently mentioned attributes included colors (5/8 participants) and background (3/8 participants). Some attributes were more specific to particular topics, such as ethnicity. For example, E8, when searching for images of a person sitting next to a machine, stated, \qquote{I want [the word] "people" to encompass all types of people. I want to specify that I want diversity in ethnicity}.

To verify image diversity, participants applied their own criteria. Six out of 8 participants considered images diverse when they covered a wide range of different labels.
When exploring communal spaces, E4 noted that she \qquote{imagined a variety, but it's pretty much the same option in terms of types of activities.} 
Similarly, E3 observed a lack of diversity in images of buildings, stating that there was \qquote{not a lot of different architectural style, not enough variety}.

\subsubsection{Methods for Controlling Diversity}

Participants diversified their images by adjusting the proportions of each label, sometimes in an imbalanced manner, to prioritize some labels over others.
While searching for a composition featuring a manhole, E5 stated that it would be better to have \qquote{most of the images with a centered composition [with the manhole in the center] and just some images with different compositions, with the manhole in one of the corners}.

Struggling with prompt-engineering, E2 obtained text suggestions to \qquote{help me improve my imagination for the prompt. Maybe I didn’t think about pink or steampunk, and I can consider them now.} Likewise, E6 wanted \qquote{suggestions of prompts, to see what images are possible to obtain and view them all at the same time}.

Finally, two participants wanted to modify multiple images as a batch. 
E1 \qquote{wanted to replace some [images] with the same [ones], but with the feet on top of the head} to diversify the orientation of a monkey, and 
E6 wamted to \qquote{have [the same frog] in different environments and settings}.

\subsection{Design Goals}
\label{designRequirements}
Building on the findings of the formative study, we propose several design goals to enable users to control image diversity through user-defined attributes to align with user specifications.
\begin{itemize}
    \item[S1)] \textbf{\Gen \space a collection of images}. 
    Users generate an initial set the images with a simple text prompt, view the images as a whole to observe their variability and assess the need for diversity.
    \item[S2)] \textbf{\Verif \space image diversity based on attributes}. 
    Users specify diversity in terms of attributes they want to see vary, and
    view the prevalence of various labels in a histogram.
    Attribute labels are initially proposed using an LLM.

    \item[S3)] \Vary \space \textbf{attribute label distribution}. 
    Users adjust the histogram to change labels or their proportions.
    These labels are sampled and appended to the original prompt as modifiers to adapt each image.

    \item \Iter \space \textbf{image generation}. 
    Users generate images based on the new attribute label distribution, and can repeat steps S1--S3.
    We also include an action history to review or revert changes.

\end{itemize}

\section{Varif.ai Interaction Design}
\label{designPresentation}

\begin{figure*}[t]
  \centering
  \includegraphics[width=0.8 \linewidth]{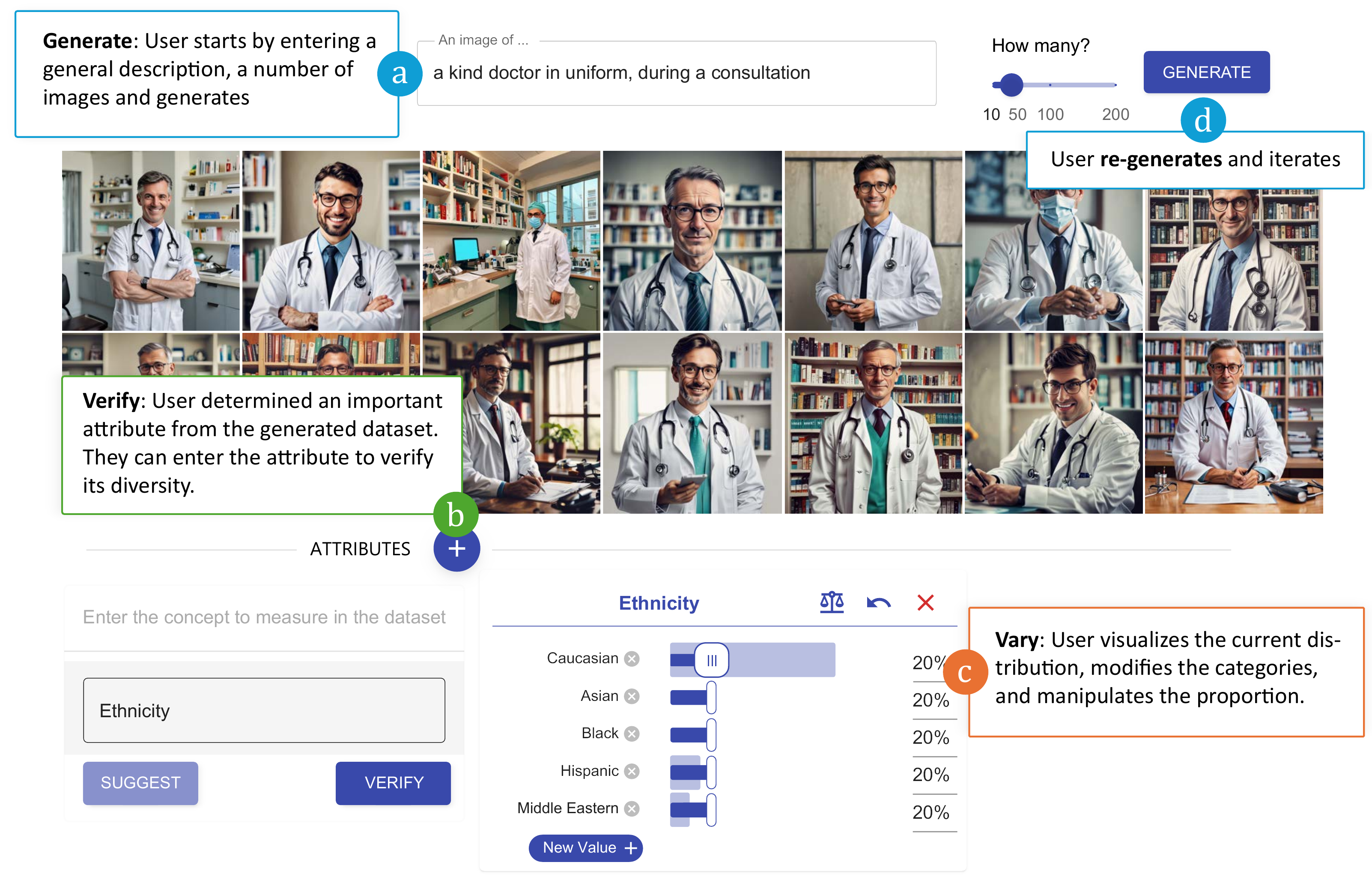}
  \caption{
  User interaction flow for the doctor scenario described in Section~\ref{designPresentation}.
  }
  \label{fig:interaction}
  \Description{The figure illustrates the Varif.ai workflow with four labeled steps: "Generate (a): Users input a prompt (e.g., "a doctor") and select the number of images to generate. The generated image grid displays various doctors with diverse appearances. Verify (b): Users identify an important attribute from the dataset (e.g., "Ethnicity") and enter it into the attributes panel to verify its diversity in the generated set. Vary (c): The system displays a distribution of the selected attribute (e.g., ethnic groups). Users can adjust the sliders to modify the categories and their proportions. (Re)Generate (d): After adjusting the attribute settings, users generate a new image set, iterating the process to achieve the desired diversity.}
\end{figure*}

Following our design goals, we developed Varif.ai (see user interface in Fig. \ref{fig:interface}) to facilitate users to verify and vary the diversity of image generation. 
Here, we describe its capabilities with a scenario to generate diverse images of doctors with various ethnicity (see interaction flow in Fig. \ref{fig:interaction}). 

\subsection{Initialization} To \Gen\space an original image collection, Varif.ai allows the user to input a context prompt\footnote{Similar to popular prompt-based interfaces~\cite{AUTOMATIC1111_Stable_Diffusion_Web_2022, brade2023promptify, feng2023promptmagician}.},
such as \qquote{an image of a kind doctor during a consultation}, specify 10 images, and click
the \raisebox{-4pt}{\includegraphics[scale=1]{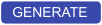} \Description{Generate}} button.

The images are generated and displayed in a gallery for the user to review. 
Clicking on each image enlarges it for close inspection.
In this scenario, the user identifies a lack of social diversity in the generated images, as most of them depicted white, old men. 

\subsection{Verify Diversity}

As the user identifies specific attributes to explore, namely ethnicity, gender and age, the user starts by adding one to the Attributes tab. After entering ``Ethnicity,'' Varif.ai suggests a set of labels generated by a large language model (LLM): 
Caucasian, Black, Asian, Hispanic, and Middle-Eastern. 
The user can modify, add, or remove any of the labels. 
Furthermore, if the user needs help, the user can click the \raisebox{-4pt}{\includegraphics[scale=1]{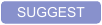} \Description{Suggest}} button, to ask the LLM (like \cite{hayati2024diversity, suh2023structured}) to provide possible attributes based on the context prompt.
Varif.ai then classifies the generated images to the labels, counts them and visualizes their distribution in a histogram for each attribute (\Verif). 
Hovering the mouse over each image shows a pop-up stating its predicted labels.
Hovering over each histogram bin highlights images with that attribute label.
In this scenario, the user examines the histogram and observes that over 80\% of the images are classified as Caucasian, with the remaining images as Hispanic and Middle-Eastern. This indicates a significant lack of diversity in the image set.

\begin{figure*}[t]
  \centering
  \includegraphics[width=13.0cm]{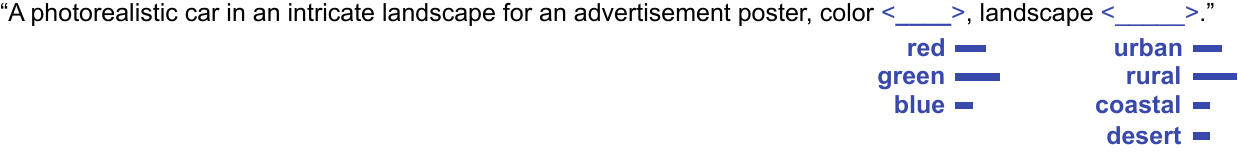}
  \caption{Conceptual diagram of probabilistic prompts. For each attribute (e.g., color), Varif.ai samples labels (e.g., red, green, blue) according to user-defined probabilities (e.g., 40\%, 50\%, 10\%) and replaces the placeholder in the context prompt. When sampled repeatedly, generated image sets reflect the specified distributions.}
  \label{fig:conceptVary}
  \Description{An image displaying a text prompt for generating images of cars:
“a photorealistic car in an intricate landscape for an advertisement poster, color <>, landscape <>.”
The prompt contains two blank fields meant to be filled in with user-specified values for the car’s color and the landscape type.
On the right, possible values are listed in two columns:
Color options: red, green, blue
Landscape options: urban, rural, coastal, desert
Each option is followed by a black horizontal bar, likely indicating a selection interface or stylistic emphasis.}
\end{figure*}

\begin{figure*}[t]
  \centering
  \includegraphics[width=0.85\linewidth]{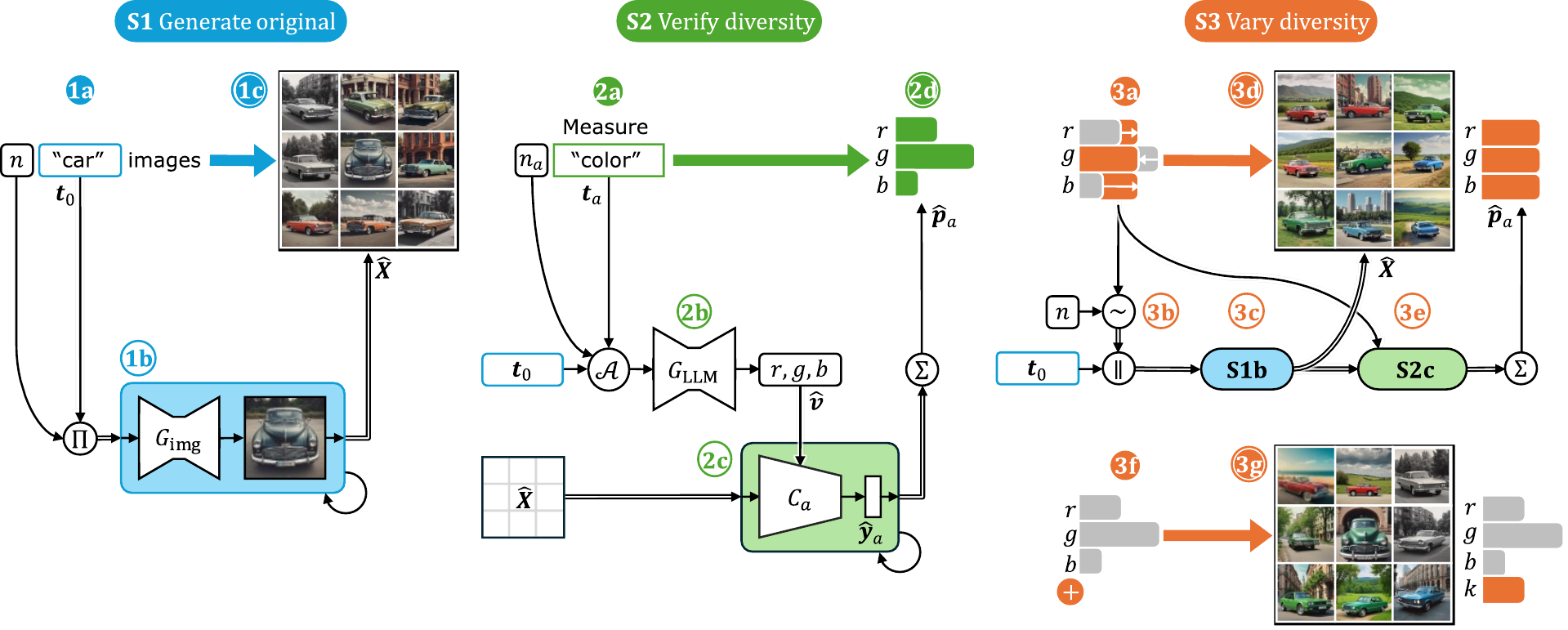}
  \caption{Varif.ai architecture. 
  S1) \Gen\space an image collection with an initial prompt $t_0$, duplicated ($\Pi$) $n$ times and fed into an image generation model $G_{img}$ to obtain generated images $\hat{X}$. 
  S2) \Verif\space attribute diversity, each from an attribute $t_a$ and the prompt $t_0$, combined in a prompt template $\mathcal{A}$ and fed into a language model $G_{LLM}$ to obtain $n_a$ labels. Each image is then labeled using a vision-language model $C_a$, and the distribution are obtained by counting ($\Sigma$) the labels in the image set. 
  S3) \Vary\space attribute diversity to generate images following the user-defined distribution, where $n$ labels are sampled ($\sim$) and appended ($\|$) to the original prompt $t_0$ to obtain extended prompts, and executing Step S1b and S2c again.}
  \label{fig:architecture}
  \Description{The figure illustrates the technical workflow of Varif.ai, broken down into three main stages: S1) Generate Original, S2) Verify Diversity, and S3) Vary Diversity. In S1 (Generate Original), the process begins by generating a set of images (shown as a grid of car images). These images are produced based on an initial input or prompt, represented by the notation "t" (for "car"), passed through a generation model. The original set of generated images is displayed on the right. In S2 (Verify Diversity), the diversity of the image set is measured based on a user-defined attribute (e.g., "color"). This attribute is analyzed, and the diversity of the attribute (e.g., red, green, blue labels) is visualized through histograms. The histograms allow users to see how balanced the attribute distribution is across the image set. In S3 (Vary Diversity), users can adjust the labels (shown as histograms) to introduce more diversity or refine the images based on user specifications. This modification process then leads to the re-generation of a new image set with more diverse attributes, which is displayed on the right side. Users can iterate between S2 (verifying diversity) and S3 (varying diversity) to fine-tune the image set.}
\end{figure*}

\subsection{Vary Diversity}
Each attribute histogram is interactive with sliders (Fig. \ref{fig:interaction}c) that the user can \Vary \space diversity to achieve goals such as increasing fairness, prioritizing specific labels, improving coverage, or removing undesired labels. 
We include a convenient Balance button \raisebox{-2pt}{\includegraphics[scale=1]{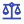}} to set the histogram to a uniform distribution 
In this scenario, the user improves coverage and fairness by adding the ``Native American'' label and by balancing all attribute labels across ethnicities. 

\subsection{Iterate}
After clicking the \raisebox{-4pt}{\includegraphics[scale=1]{figures/generate_button.pdf} \Description{generate}} button again, Varif.ai generates a new set of images aligned to the user's specified distribution.
The histograms and pop-up tooltips are updated to reflect the new distributions and labels. 
The user can track changes and branch from any previous iteration using the Message History bar.
Examining the updated histogram, the user verifies that Varif.ai has successfully diversified ethnicities. 
Subsequently, the user can \Iter\space to verify and vary the diversity of other attributes---gender and age.

\section{Technical Approach}
\label{sec:implementation}

We describe our technical approach to support the design features of Varif.ai to: 
S1) generate image sets, either initially or iteratively, using a text-to-image diffusion model, 
S2) verify diversity through image-to-text modeling, and
S3) vary diversity by sampling the text prompt distribution based on the user-defined attribute histograms, and reapplying steps S1 and S2.
Architecture shown in Fig. \ref{fig:architecture}.

\subsection{Generate Original} 
\label{tech:generate}
Given a description prompt provided by the user, the first step is to generate an image collection using a diffusion model.
\textcolor{BLUE}{1a)} The user defines a text prompt~$t_0$ that describes the common theme of the images, and the number of desired images~$n$ in the user interface. 
\textcolor{BLUE}{1b)} $t_0$ is duplicated $n$~times and each prompt is input to a text-to-image generation model~$G_{\text{img}}$. 
Since our method only modifies the text prompts, any text-to-image generation model can be used. 
\textcolor{BLUE}{1c)} The generated images~$\hat{\bm{X}}$ are displayed in the user interface.

\subsection{Verify Diversity}
To determine diversity along an attribute, multiple labels for the attribute must be measured.
Since the attribute~$t_a$ is open-ended, Varif.ai uses an open-world large-language model (LLM) to propose $n_a$ labels. 
\textcolor{GREEN}{2a)} 
The attribute~$t_a$, the context~$t_0$, and the number of labels~$n_a$ are combined in a prompt template~$\mathcal{A}$ (see Appendix~\ref{app:llm}), and
\textcolor{GREEN}{2b)} input to LLM~$G_{\text{LLM}}$ to generate attribute-label prompts~$\hat{v}$ (examples in Table~\ref{tbl:scales}). 
If users need attribute suggestions, Varif.ai can input the context~$t_0$ to $G_{\text{LLM}}$ to generate a suggested attribute~$t_a$.

\begin{table}[t]
\centering

\caption{Example labels proposed by $G_{LLM}$ for the same attribute and different contexts.}

\begin{tabular}{lp{18.0em}}
\hline
 \addlinespace[0.05cm]
 Context $\bm{t}_0$& Generated labels $\hat{\bm{v}}$ for attribute $\bm{t}_a$ \qquote{age} \\ 
\addlinespace[0.05cm]
\hline 
\addlinespace[0.05cm]
 Person & Child, Adolescent, Young Adult, Middle-Aged, Elderly \\
  \addlinespace[0.05cm]
 Doctor & 30s, 40s, 50s, 60s, 70s  \\
 \addlinespace[0.05cm]
 Bridge &   Newly built, Recent, Old, Ancient, Historic   \\
 \addlinespace[0.05cm]
 Car & New, Used, Old, Classic, Vintage \\
 \addlinespace[0.05cm]

 \hline
\end{tabular}
\Description{"The table presents labels proposed by a generative model for the attribute  age  in different contexts. The table adapts the concept of  age  based on the specific context (e.g., person, doctor, bridge, car). The columns represent four different contexts--- Person,   Doctor,   Bridge,  and  Car ---and the rows show how the model adapts the attribute  age  to each context. For Person: The age groups are Child, Adolescent, Young Adult, Middle-Aged, and Elderly. For Doctor: The ages are represented as 30s, 40s, 50s, 60s, and 70s. For Bridge: The age is expressed as Newly built, Recent, Old, Ancient, and Historic. For Car: The age is described as New, Used, Old, Classic, and Vintage. The table illustrates how the generative model adapts the age scale to match the context, demonstrating flexible interpretation across different domains."}
\label{tbl:scales}
\end{table}

For each attribute, an image $\hat{\bm{x}} \in \hat{\bm{X}}$ is labeled via classification using CLIP~\cite{radford2021learning}, since we are using user-defined, open-domain class labels instead of predetermined fixed classes with pretrained classifiers.
\textcolor{GREEN}{2c)} For each attribute label, we compute the CLIP similarity score between the image~$\hat{\bm{x}}$ and label~$\hat{v}$, then choose the label with the highest similarity $\hat{y}_a$. This is the predicted label for the image.
\textcolor{GREEN}{2d)} We then count the labels in the image set to determine the distribution $\hat{p}_a$, which is visualized in the attribute histogram.

\subsection{Vary Diversity} 

To vary image diversity based on attributes, 
\textcolor{ORANGE}{3a)} the user can adjust the sliders in the histogram, which modifies the distributions~$\hat{\bm{p}}_a$.
Varif.ai generates prompts \textit{probabilistically} by 
\textcolor{ORANGE}{3b)} sampling $n$ labels from the distribution of attribute labels (see Fig.~\ref{fig:conceptVary}) and appending to the original context prompt~$\bm{t}_0$, and \textcolor{ORANGE}{3c)} reapplying step S1b to \textcolor{ORANGE}{3d)} generate a new image set to better align with the new distribution.
\textcolor{ORANGE}{3e)} Step S2c is reapplied to verify the distributions and update the histograms.
\textcolor{ORANGE}{3f)} The user can also vary diversity by adding labels in $\hat{\bm{v}}$ to
\textcolor{ORANGE}{3g)} generate images representing the new labels.

\subsection{Implementation Details}
We implemented Varif.ai as a responsive web application. 
The front-end is implemented with Material UI components\footnote{\href{https://mui.com}{mui.com}}, 
the back-end server with Python Flask, and
all machine learning models with Pytorch\footnote{\href{https://pytorch.org}{pytorch.org}} and Diffusers\footnote{\href{https://github.com/huggingface/diffusers}{github.com/huggingface/diffusers}}. 
For specific models, we used the open-source SD-XL Lightning~\cite{lin2024sdxl} for text-to-image generation, and the open-weight LLM LLaMA-2~\cite{touvron2023llama} to generate suggestions.
We used 8-bit quantized versions of models to accelerate inferences. 
To ensure the LLM returns lists of labels, we employed the guidance library~\footnote{\href{https://github.com/guidance-ai/guidance}{github.com/guidance-ai/guidance}}, with the specific instructions provided in Appendix~\ref{app:llm}.
Varif.ai ran on an Nvidia H100 GPU, and generating 10 images takes around 6 seconds. Our code is available at: \githublink

\begin{figure*}[h]
  \centering
  \includegraphics[width=\linewidth]{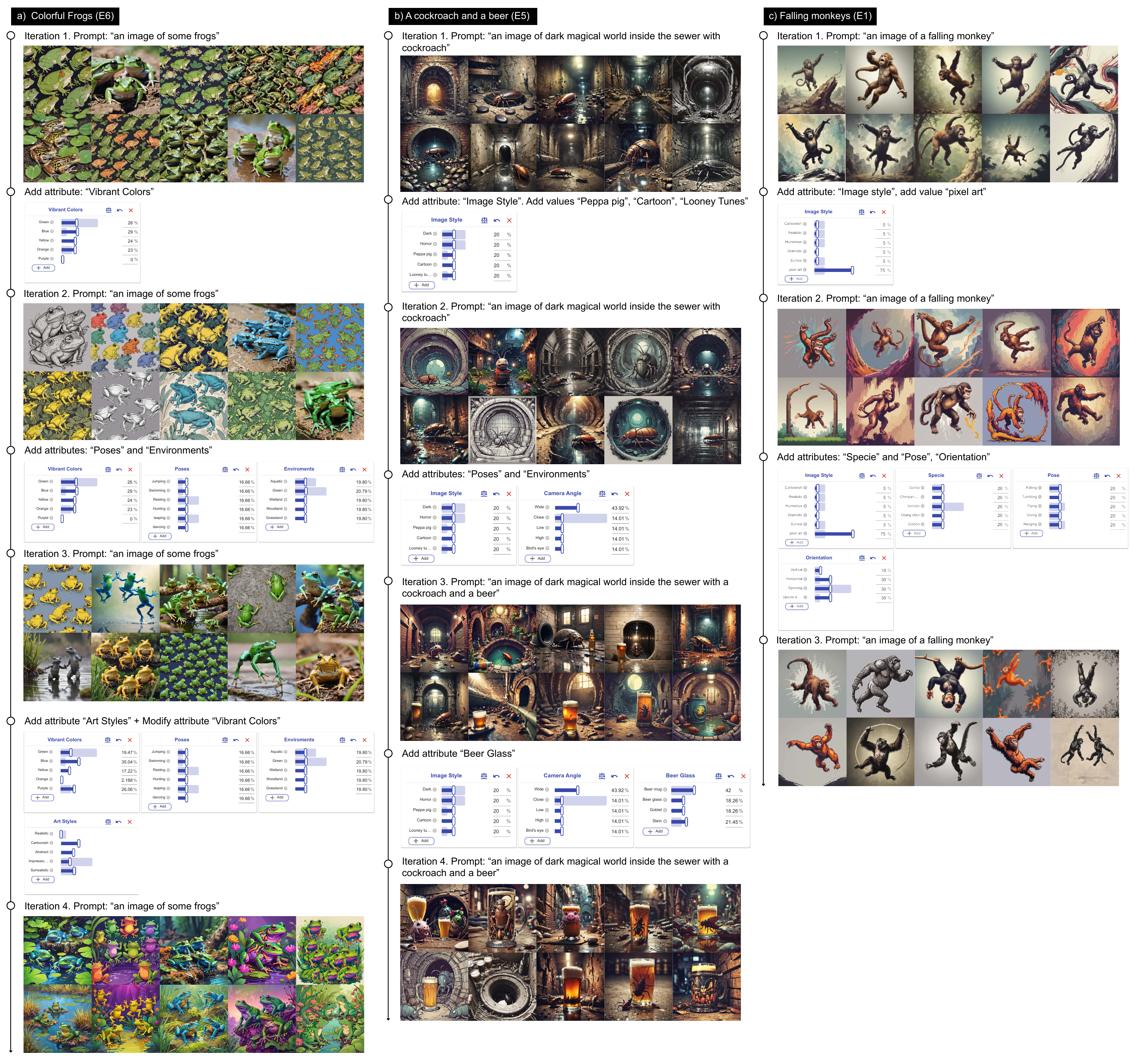}
  \vspace{-0.3cm}
  \caption{Examples of participants iteratively adding and adjusting attributes distributions, and revising the context prompt.
}
  \label{fig:iterations}
  \Description{This figure illustrates how participants iteratively diversified images by adding or modifying attributes across four scenarios: (a) Falling Monkeys, (b) A Cockroach and a Beer, (c) Colorful Frogs, and (d) Custom Art Styles. For each scenario, the iterations are displayed vertically, showing the progression of prompts and attributes over time.
Key elements include:
Attributes: Examples like "Image Style," "Pose," "Orientation," and "Environment" are added or modified to refine image generation.
Iteration Details: Each row represents an iteration with changes to the attributes, prompts, or values applied.
Example topics: Specific prompts, such as "an image of a falling monkey" or "an image of some frogs," evolve with new attribute values like "Pixel Art" or "Vibrant Colors."
Diversity Metrics: Attribute adjustments, such as adding or balancing percentages, are noted alongside changes in the generated images.
The figure demonstrates how participants explored iterative adjustments to control image diversity and achieve their desired outcomes effectively.
}
\end{figure*}

\section{Formative User Study}
\label{sec:formativeStudy}

We conducted an elicitation study to examine whether Varif.ai satisfies our design goals from Section~\ref{sec:elicitationStudy} and determine its usability.

\subsection{Participants and Study Procedure}

We re-invited the same 8 participants from the formative study to use Varif.ai using the same or similar scenario that they had explored previously. Each session lasted 30 minutes on Zoom with recorded screensharing and using the think-aloud protocol.
At the end of the session, we conducted a semi-structured interview, focusing on Varif.ai's usability and the helpfulness of its components.

\subsection{Findings}

We performed a thematic analysis and affinity diagramming~\cite{lucero2015using} of participant usage and utterances.
We report the usage patterns of Varif.ai, its perceived usefulness, weaknesses, and feature requests.

\subsubsection{Usage Patterns of Varif.ai}

Participants exhibited different usage patterns (Fig.~\ref{fig:iterations}).
Six out of 8 participants iteratively added attributes to explore \pquote{with and without the attribute}{E3} and test if \pquote{adding one [attribute] doesn't mess everything up}{E4}. 
Participants iteratively adjusted the attribute distributions while monitoring if changes were effective or non-responsive.
For example, wanting images of colorful frogs, E6 first measured the attribute ``vibrant colors'' (Fig.~\ref{fig:iterations}a, Iteration 1) and observed too many green frogs. 
Despite balancing the proportion of \qquote{this category [...] because there are still too many [green frogs]} (Iteration 2).
She overcompensated to further reduce green frogs by setting the label a minority.

In addition to interacting with the attribute histograms, some participants modified the original context prompt.
For example, E5, an artist searching for images of cockroaches with a beer, initially adjusted attributes to generate images without beer (see Fig.~\ref{fig:iterations}b, Iterations 1-2). She then modified the prompt to include beer (Iterations 3-4), explaining that she \qquote{wanted to see how the system would do without the beer first, and then add something more complex}.

\subsubsection{Feedback on Feature Helpfulness}

Participants generally found Varif.ai's components helpful, and readily used labels and histograms. 

\textit{Attributes}.
Participants appreciated the ease of use of the Attributes tab in Varif.ai. 
E6 noted, \qquote{once I understood how it works it’s a lot easier to explore what I want. It's so much easier to type [the attribute] instead of typing [each label] one by one}. 
However, two participants reported ambiguity when using intricate prompts or many attributes, as it was unclear to which object an attribute affected. 
For example, E6 observed that the attribute ``Vibrant colors'' applied to the background instead of the frogs (see Fig.~\ref{fig:iterations}a, iteration 4). 
Similarly, E3 \qquote{want[ed] the color for the building, not the background, [...] I'll just try to add "building color"}, which ended up properly targeting buildings. 

\textit{Attributes Labels}.
Participants appreciated both the label suggestions and editability. While looking for community centers, 
E4 found the suggested labels for ``Crowd'' (Crowded, Empty, Busy, Vibrant, Quiet) \qquote{very smart, I would have never thought about all of them}. 
Yet, participants may want more than what was suggested. While looking for cockroach references, E5 added unconventional labels under the ``Image Style'' attribute, such as ``Peppa Pig'' or ``Looney Tunes'' (see Fig.~\ref{fig:iterations}b, iteration 1) to \qquote{see if it gives a more cartoony and childish vibe} to the images.
She liked that label editing \qquote{helps you to really customize and add your own vision inside. What [generated images] depict is very common, and so it helps getting inspired to put something completely off}.

\textit{Distribution of Attribute Labels}.
Participants appreciated being able to control the proportions of attribute labels using the sliders. 
Three participants deliberately assigned low probabilities to labels they \pquote{don’t want, [but still] prefer to have some examples to compare with what I want}{E3}.
While exploring images of falling monkeys, E1 strongly preferred a pixel-art style but still included other styles \qquote{just to keep options open a little bit} (Fig.~\ref{fig:iterations}c, Iteration 1). 

\subsubsection{Usability issues}

Participants noted some weaknesses due to technical issues in the underlying models affecting the UI behavior.

\textit{Inaccurate Attribute Label Distributions.}
Some participants found that the measured attribute histograms were inconsistent with the actual labels of the generated images. 
For example, E6 observed that frogs were consistently classified as green, despite most not appearing so (see Fig.~\ref{fig:iterations}a, iteration 4). 
She speculated, \qquote{I don’t know why it keeps saying that most frogs are green. Maybe it’s because of the grass,} highlighting the challenge of accurately attributing labels to specific objects within an image. 

\textit{Entangled Attributes}.
Two participants found it limiting that attributes were treated independently, and wanted to conjoin them. 
For instance, E4 sought examples of communal spaces that were \qquote{crowded by day but empty at night}, but could not achieve this by adding the attributes ``Crowd'' and ``Time of the Day.''
Conversely, latent interactions between attributes occasionally led to unintended but welcomed diversity. 
While searching for an image of friends in Scotland for an ad campaign, E2 observed that diversifying the attribute ``Weather'' resulted in \qquote{a lot more diversity [than at the beginning], not just in weather but also in landscapes.}

\subsubsection{Feature Requests}

Several participants requested additional features. Artist E1, who had a distinctive personal style, expressed interest in using custom \textit{seed images}, asking whether it was possible \qquote{to put my own images in the interface, and try from there.}
E2, encountering undesirable outcomes, wanted to \textit{suppress or prevent certain attribute labels} \qquote{to remove completely the buildings.}

\section{Summative user study}
\label{sec:summativeStudy}

\begin{table*}[t]
  \caption{Criteria for the three diversity specification degrees (columns) and the three scenarios (rows.)}
  \vspace{-0.3cm}
  \Description{The table presents values of diversity specification degrees across three scenarios--- Doctors, Birds, and Cars ---with three specification degree:  Open-ended ,  Attributes , and  Distributions. -  Scenario: Doctors  -  Open-ended : You are a kid book illustrator wanting to generate inspiration images for a kid book, portraying some doctors.  -  Attributes : Increase diversity in gender, race, environment, and image style.   -  Distributions : Satisfy the following: 75\% women, 25\% men, 33\% Black, 33\% Asian, 33\% White. 50\% hospital, 25\% home consultation, 25\% office. 70\% photos, 30\% cartoon. -  Scenario: Birds  -  Open-ended : You are an elementary teacher wanting to show your students the variety of birds that exist.  -  Attributes : Increase diversity in type of birds, habitat, and pose.  -  Distributions : Satisfy the following: 100\% photos, 40\% forests, 10\% each of savannah, desert, polar, swamp, river, and sea. 25\% each of flying, on a branch, on the ground, or on water. 10\% tall birds, 70\% small birds, 20\% others. -  Scenario: Cars  -  Open-ended : You are a car enthusiast wanting to see cool references for a car poster/wallpaper.  -  Attributes : Increase diversity in color, environment, and weather conditions.   -  Distributions : Satisfy the following: 100\% photorealistic images, 20\% each of blue, red, yellow, green, purple. 40\% sunny, 20\% each of snowy, cloudy, or rainy. 75\% in the city, 25\% in the countryside. The table illustrates how the criteria for generating images become more specific as they progress from open-ended descriptions to detailed distributions.}
  \label{tab:condition}
  \begin{tabular}{lp{12em}p{12em}p{24em}}
    \toprule
    Scenario & Open-ended & + Attribute-specific & + Attribute-label-specific \\
    \midrule
     & ``You are ...  & ``Increase diversity in ... & ``Satisfy the following ... \\
    \addlinespace[0.05cm]
    \addlinespace[0.1cm]
    \textit{Doctors} 
    & ``a kid book illustrator wanting to generate inspiration images for a kid book, portraying some doctors.''
    & ``gender, race, environment, and image style.''
    & ``75\% women and 25\% men, have 33\% Black, 33\% Asian and 33\% White people. Have 50\% hospital, 25\% home consultation and 25\% office. Have 70\% photos and 30\% cartoon.'' \\
    \addlinespace[0.05cm]
    \addlinespace[0.1cm]
    \textit{Birds} 
    & ``an elementary teacher, you want to show your students the variety of birds that exist.''
    &  ``type of birds, habitat, and pose.''
    & ``100\% photos, Have 40\% forests, and 10\% each: savannah, desert, polar, swamp, river and sea. Have 25\% each: flying, on a branch, on the ground, or on water. Have 10\% tall, small birds, 70\% others.''
    \\
    \addlinespace[0.05cm]
    \addlinespace[0.1cm]
    \textit{Cars} & ``a car amateur wanting to see cool references for a car poster/wallpaper.''
    & ``color, environment, and weather conditions.''
    & ``100\% photorealistic images. Have 20\% each: blue, red, yellow, green, or purple. Have 40\% sunny, and 20\% each: snowy, cloudy, or rainy. Have 75\% in the city, 25\% in the countryside.'' \\
    \addlinespace[0.05cm]
  \bottomrule
\end{tabular}
\end{table*}

Having found that users can use Varif.ai in our formative user study, we next conducted a controlled study to investigate whether Varif.ai improves \textit{image diversity} compared to the baseline prompt engineering, and under what settings. 
Furthermore, our earlier elicitation study found that using prompt engineering was tedious, 
so we also examined whether Varif.ai improves \textit{user engagement} in terms of time engaging on the task and number of iterations.

For external validity, we note that users may want to specify diversity with different degrees: from only desiring some variety, to defining specific attributes to diversify, to prioritizing specific attribute labels or requiring equal representation across labels.
Specific to Varif.ai, we evaluated the diversity of attributes conceived by users, \textit{alignment} to the target distribution of the aforementioned diversity degrees, and user effort to modify and finetune attributes.

We focus on simpler prompt tasks with basic scenarios to 
limit overfitting to contexts.
Our research questions are:
\begin{enumerate}[label=RQ\arabic*.]
    \item Does Varif.ai improve a) image diversity and b) user engagement compared to prompt engineering?
    \item How are a) image diversity, b) attribute count, c) diversity alignment, and d) user engagement affected by the degree of diversity specification?
\end{enumerate}

\subsection{Experiment Method}

\subsubsection{Experiment Design}

We varied Technique and Diversity Specificity Degree with Scenario as random variable, and 
measured Image diversity, Task time, Iteration count, and Attribute count.

\begin{figure}[t]
  \centering
  \includegraphics[width=0.98\linewidth]{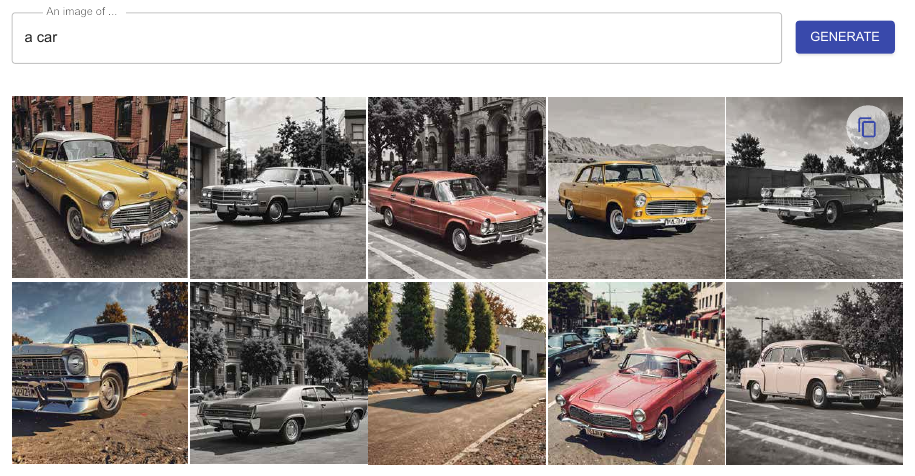}
  \vspace{-0.2cm}
  \caption{
    Prompt-only UI without diversity controls.
    }
  \label{fig:baseline}
  \vspace{-0.2cm}
  \Description{This figure represents the baseline user interface for generating images. The interface includes a simple text input field where users can describe their desired image (e.g., "a car") and a "Generate" button to initiate image creation. The design highlights a minimalistic approach, focusing on ease of use with straightforward prompt-based input and image generation functionality.}
\end{figure}

\textit{Technique (IV1)}.
We compared Varif.ai to Prompt-only (see Fig.\ref{fig:baseline}), which is an ablated version of Varif.ai with only the prompt text entry and no Attributes tab to view or add attributes. 
This was done within-subjects with counterbalancing to mitigate learning effects.
We also compare against the original images generated from the initial prompt, without participant interaction, to analyze how much Prompt-only and Varif.ai increased diversity.

\textit{Diversity Specificity Degree (IV2)}.
We examine how different degrees of specificity affect participants' use of Varif.ai. 
\textit{Open-ended} provides minimal guidance, offering a one-sentence scenario to explain the need for diversity and engage participants~\cite{kujala2002user}.  
\textit{Attribute-specific} defines several attributes for participants to focus on when increasing diversity in addition to the open-ended criteria.
\textit{Attribute-label-specific} defines precise attribute labels and their distributions that participants are asked to satisfy in addition to the attribute-specific criteria.
To avoid biasing participants in the open-ended condition, we present the degrees sequentially, from least to most detailed.
To avoid overwhelming participants with a $3 \times 2$ condition setup, we only tested the Prompt-only interface with the open-ended specification degree, since we learned from our formative study that it would be too tedious for users to specify at scale. 

\textit{Scenarios (RV)}.
We use three scenarios---\textit{Doctors}, \textit{Birds}, and \textit{Cars}---treated as a random variable between-subjects. These scenarios require no prior knowledge and are common enough to be generated well by diffusion models~\cite{samuel2024generating}.
We include a social diversity scenario (\textit{Doctors}) due to biases in generated images toward specific populations~\cite{basu2023inspecting}, and 
scenarios with animate subjects (\textit{Birds}) and inanimate objects (\textit{Cars}) to explore potential differences in how participants approach diversity. Table~\ref{tab:condition} summarizes the criteria provided for each diversity specification degree and scenario.

\begin{figure*}[t]
  \centering
  \includegraphics[width=0.6\linewidth]{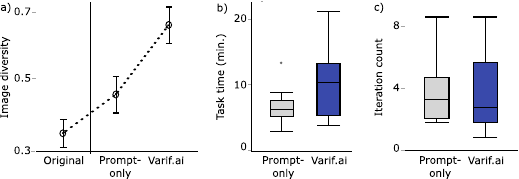}
  \caption{Results of a) Image diversity, b) Task completion, and c) Iteration count with different techniques. Dotted lines indicate extremely significant comparisons ($p<.0001$); otherwise, the $p$-value stated when comparisons are very significant. Solid lines indicate no significance ($p>.01$). Error bars represent a 90\% confidence interval. Black vertical line separates the theoretical baseline (no users) from the user results.}
  \label{fig:rq1}
  \Description{Three-part figure comparing Prompt-only and Varif.ai conditions.
a) Box plot of task time (in minutes) shows longer median task time and greater variability for Varif.ai than Prompt-only.
b) Box plot of iteration count shows slightly higher and more variable iterations for Varif.ai.
c) Line chart with error bars showing image diversity scores: Original (lowest), Prompt-only (moderate), and Varif.ai (highest), with a clear upward trend across conditions.
}
\end{figure*}

\textit{Computed Image Diversity (DV)}.
We collect all the images generated by the users to measure their diversity. We augment each participant's image collection to 50 total images with the original attribute specification. This increased sample size allows us to compute diversity more precisely.
From our elicitation study, participants generally considered diversity in terms of coverage. We thus measure how widely the concepts in the images are spread out as a proxy of diversity. 
We use the diversity \textit{span} metric~\cite{cox2021directed,brown2015multi}, which represents the "radius" of the images' distributions to estimate coverage. More precisely, given image embedding~$\bm{z}$ from the semantic image feature encoder CLIP~\cite{radford2021learning}, 
$\text{span}(\bm{z}) = \text{percentile}_{95\%} \|\bm{z} - \bar{\bm{z}}\|$. 
For a fair evaluation, similar to \cite{shen2023finetuning}, we use a more powerful CLIP model\footnote{\href{https://huggingface.co/laion/CLIP-ViT-bigG-14-laion2B-39B-b160k}{huggingface.co/laion/CLIP-ViT-bigG-14-laion2B-39B-b160k}} than the one used in Varif.ai.

\textit{Attribute Count (DV)}.
We measure the attribute counts and added labels from participants. Higher numbers suggest higher diversity for the open-ended degree, but the attributes may be extraneous for more specific diversity specification degrees.

\textit{Diversity Alignment (DV)}. To assess whether participants could satisfy precise distributions in the Attribute-label-specific diversity degree, we measured the alignment between the target distribution (Table~\ref{tab:condition}, right column) and the participant's attribute distribution. To quantify the distance between two probability distributions, we use the Kullback-Leibler divergence (KLD) and compute its inverse as \textit{Diversity alignment}. 
Lower KLD indicates more alignment.

\textit{Task Completion Time (DV)} and \textit{Iteration Count (DV)} were used to measure user engagement. Since we are not focusing on efficient image generation, and creative ideation requires reflection, we consider the time taken to generate images as indicative of engagement. 
We expect more inspired participants to ideate for longer, and frustrated participants to engage less.

\textit{Attribute Label Modifications (\%) (DV)} measures the effectiveness of the LLM in providing aligned labels. We considered it tedious if users need to modify a large amount of labels manually. We expect users to modify labels most often for the Attribute-label-specific diversity degree due to its stringent requirements.

\subsubsection{Procedure}

Each participant went through the following:
\begin{enumerate}[label=\arabic*)]

\item \textit{Introduction (10 min)} to the research background, including the motivation and study protocol. Demographic information was collected, and participants provided consent to record their operations and results for analysis.

\item \textit{Tutorial (5 min)}
on the main interactions, explaining their purpose and functionality.

\item \textit{Main Session (40 min)}. 
Randomly assigned a scenario task to generate images.
\begin{enumerate}[label=\alph*)]
    \item Generate images iteratively for the Open-ended degree of specification using Prompt-only, then Varif.ai (or in reversed depending on random assignment).
    \item Generate images iteratively for the Attribute-specific degree for Varif.ai only.
    \item Generate images iteratively for the Attribute-label-specific degree for Varif.ai only.
\end{enumerate}

\item \textit{Free Session (Optional)} to use Varif.ai on a scenario of their choice. Eight participants explored on their own briefly.
\end{enumerate}

\subsection{Analysis and Results} 
We recruited 20 participants through a local university mailing list and mutual connections. Participants ranged in age from 21 to 33 (Median = 23), were 12 female, and had diverse educational backgrounds: social sciences (6), medicine (3), arts (2), business (3), and STEM (6). 
Three participants reported some experience in image creation.
Nine participants had no prior experience with text-to-image models, while the others had only used them a few times. 
No one reported regular or professional use of text-to-image models.
The experiment, conducted over Zoom, lasted one hour with \$15.30 USD in local currency as compensation.
Participants shared their screens and were asked to think aloud. 

For statistical analysis, we fit linear mixed-effects models for each dependent variable (DV) and performed post-hoc contrast tests for specific differences that we highlight. 
Next, we discuss significant results (p < .01).

\begin{figure}[t]
  \centering
  \includegraphics[width=\linewidth]{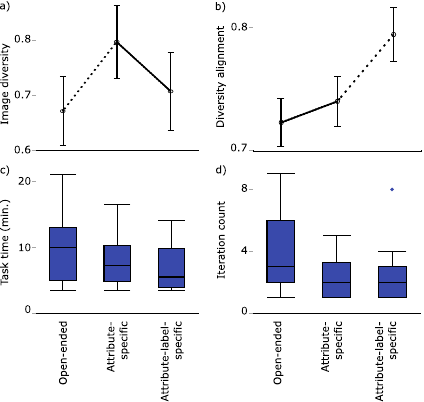}
  \caption{Results of a) Image diversity b) Diversity alignment c) Task completion and d) Iteration count for different specificity degrees. Dotted lines indicate extremely significant comparisons ($p<.001$); otherwise, the $p$-value stated when comparisons are very significant. Solid lines indicate no significance ($p>.05$), and error bars 90\% confidence interval.}
  \label{fig:rq2}
  \Description{"Four subplots comparing user task performance and diversity outcomes across three conditions: open-ended, attribute-specific, and attribute-label-specific. (a) Boxplot showing task time in minutes, decreasing slightly across conditions. (b) Boxplot showing iteration count, with a notable outlier in the open-ended condition. (c) Line plot with error bars showing image diversity, peaking in the attribute-specific condition. (d) Line plot with error bars showing diversity alignment, increasing steadily across conditions."

}
\end{figure}

\subsubsection{Varif.ai Improved Diversity and Engagement Compared to Prompt-Only (RQ1)}

Participants unanimously found Varif.ai better than Prompt-only to diversify images with an open-ended goal.

\textit{Computed Image Diversity (RQ1a).} Varif.ai significantly improved diversity (span mean $M=0.65$) compared to Prompt-only ($M=0.45$; p < .0001) and compared to Original ($M=0.35$; p < .0001); see Fig.~\ref{fig:rq1}a. 
Participants were more satisfied with the diversity they accomplished with Varif.ai than with Prompt-only.
P2, while diversifying images of birds using Varif.ai, commented that  \qquote{[the images] are a lot more diverse than at the beginning, there is a lot more variety of birds and backgrounds}.
There was no significant difference across scenarios, indicating the generality of the effect.

\textit{Task Time (RQ1b).} Participants were engaged more when using Varif.ai compared to Prompt-only (M = 10.2 vs. 6.3 min, p = .019); see Fig.~\ref{fig:rq1}b.
Indeed, some participants gave up earlier trying with Prompt-only, e.g., P16 \qquote{[didn't] think the AI understood what I wanted to do} after generating images of crowds of diverse doctors instead of one doctor per image, P2 \qquote{[didn't] know how to make it clearer; I don't know why it keeps giving me so many [doctors]}.

\textit{Iteration Count (RQ1b).} There was no significant difference in number of iterations whether using Varif.ai (M = 3.7) or Prompt-only (M = 3.5); see Fig.\ref{fig:rq1}b. Participants iterated as many times on Varif.ai (min=1, M=3, max=9) as on Prompt-only (min=2, M=3.5, max=9); see Fig.\ref{fig:rq1}c. 
See examples of iterative prompt refinements from the Prompt-only trials in Appendix~\ref{app:iterations}. 

\begin{table}[t]
    \caption{Attribute count and label modifications across degrees of diversity specificity. 
    Mean (Standard Deviation).
    }
    \Description{Table showing attribute count and percentage of attribute label modifications across three conditions of image diversity specificity: Open-ended, Attribute-specific, and Attribute-label-specific. Attribute counts are similar across conditions (~3.6–3.9), while label modifications increase notably in the Attribute-label-specific condition (42.7\%) compared to Open-ended (14.0\%) and Attribute-specific (13.9\%). Standard deviations are provided in parentheses.
    }
    \label{tab:llmValues}
    \begin{tabular}{p{8em}p{4em}p{4em}p{6em}}
        \toprule
        \addlinespace[0.05cm]
          & Open-ended & Attribute-specific & Attribute-label-specific \\
        \addlinespace[0.05cm]
        \hline
        \addlinespace[0.05cm]
        Attribute count & 3.86 (1.46) & 3.64 (0.49) & 3.70 (0.75)\\
        \addlinespace[0.05cm]
        Attr. label mods. (\%) & 14.0 (22) & 13.9 (26) & 42.7 (28) \\
        \addlinespace[0.05cm]

        \bottomrule
    \end{tabular}
\end{table}

\subsubsection{Varif.ai Improves Diversity and Control for More Specific Degrees of Diversity Goals (RQ2)}
Participants could better diversify images when attributes were provided, and participants could align diversity to precise criteria using Varif.ai.

\textit{Computed Image Diversity (RQ2a).} 
When given the goal of diversifying images with Attribute-specific degree, participants generated more diverse images than with Open-ended degree (span M = 0.80 vs. M = 0.66, p = .009); see Fig.\ref{fig:rq2}a.
Diversity span was not significantly higher for Attribute-label-specific degree (M = 0.71), as non-uniform requirements did not necessarily lead to high diversity outcomes.
The results generalized across scenarios, having no significant difference among them.
On their own, some participants initially struggled to identify attributes to diversify with Open-ended degree, such as P11, who wanted, \qquote{cars, trucks, sports cars, but I don't know how to say that}. However, participants were more inspired when guided with specific attributes to consider. For example, P16 \qquote{didn't think about changing the background of the doctors, that's a good idea}.

\begin{figure}
    \centering
    \includegraphics[width=\linewidth]{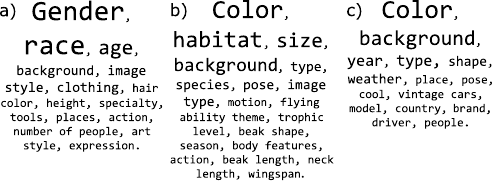}
    \caption{Word cloud of attributes defined by participants for the three scenarios a) Doctors, b) Birds, c) Cars and for the Open-ended degree. The font size is proportional to the number of participants who specified the attribute.}
    \label{fig:attributes}
    \Description{ three word clouds of participant-defined attributes for scenarios: a) Doctors, b) Birds, and c) Cars, under the Open-ended condition. In each cloud, larger font size indicates higher frequency of mention. Prominent attributes include Gender and Race for Doctors; Color, Habitat, and Size for Birds; and Color, Background, and Year for Cars.}
\end{figure}

\begin{figure*}[t]
  \centering
  \includegraphics[width=14.5cm]{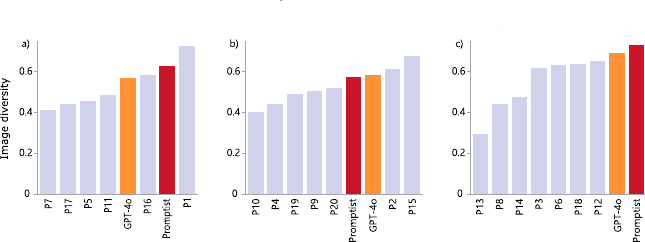}
  \caption{
  Results of image diversity from the summative user study using Varif.ai for one image set of 20 images in the Open-ended diversity degree (blue), 
  compared to automatic diversification by 
  Promptist~\cite{hao2024optimizing} (red) and 
  GPT-4o~\cite{achiam2023gpt} (orange) across scenarios a) Doctors, b) Birds, and c) Cars.
  Each bar is a single measurement, so there are no error bars.}
  \label{fig:promptist}
  \Description{Bar chart comparing image diversity scores for one image set of 20 images across three scenarios: a) Doctors, b) Birds, and c) Cars. Each subplot shows results from users (blue bars), GPT-4o (orange), and Promptist (red). Automated methods generally achieve higher diversity scores than most users. Bars are labeled with participant or method names; no error bars are shown.
}
\end{figure*}

\textit{Attribute Count (RQ2b).} Though not significantly different, participants diversified M = 3.7 attributes for each specification degree, with values more varied for the Open-ended degree ($M = 3.86$, min~= 1, max = 7) compared to the Attribute-specific degree ($M = 3.64$, min = 3, max = 4) and the Attribute-label-specific degree ($M = 3.70$, min = 2, max = 4); see Table~\ref{tab:llmValues}, first row. 
Interestingly, beyond a few attributes commonly chosen by most participants, the selection of attributes varied widely. The attributes selected with the Open-ended degree and for each scenario are shown in Fig.~\ref{fig:attributes}.
A few participants (P7, P18, P20) used the open-ended nature of attributes to add scenario-specific details. P7 included \qquote{toys [to] help with the child book}.
Participants in the \textit{Doctors} scenario primarily focused on social diversity. P1 and P5 actively reduced the proportion of labels representing stereotypical images, increasing representation for minorities to \pquote{compensate for the stereotypical images}{P5}. Additionally, participants identified more attributes to diversify animate \textit{Birds} compared to inanimate \textit{Cars}.

\textit{Diversity Alignment (RQ2c).} When prompted with the Attribute-label-specific degree, participants generated image sets that more closely matched the target distributions compared to the Attribute-specific degree (alignment $a=0.79$ vs. $a=0.74$, $p<0.001$) and the Open-ended degree ($a=0.79$ vs. $a=0.73$, p < 0.001). 

\textit{Task Time (RQ2d).}
Using Varif.ai, participants were engaged for a longer time during Open-ended compared to Attribute-specific (M = 10.2 vs. 8.2 min, p = n.s.) and Attribute-label-specific (M = 10.2 vs. 6.9 min, p = n.s.); see Fig.~\ref{fig:rq2}c. Although the difference is not significant, this task time decreased as the specificity degree increased, which can possibly be influenced by the think-aloud protocol, where participants tended to discuss the Open-ended degree more extensively, and by the practice effect of having more experience in earlier sessions and higher-degree specifications coming later. Iterations followed a similar trend, because participants explored more for the Open-ended degree.

\textit{Attribute Label Modification (RQ2d).}
Participants retained the attribute labels suggested by Varif.ai in 
$86\%$ of the cases for Open-ended and Attribute-specific, and significantly less for Attribute-label-specific with 57.3\% ($p = 0.006$); see Table~\ref{tab:llmValues}, second row. Participants modified labels the most for Attribute-label-specific to match the labels required in the criteria instructions. 
However, this can be tedious.
P20 struggled to get the right labels when adding the attribute ``type'' in the \textit{Birds} scenario with Attribute-label-specific degree, and \qquote{finally [gave] up and enter[ed] them one by one}. 

\subsubsection{Comparison with Automatic Diversified Prompts}

Our user study results had shown that using Varif.ai, participants can improve diversity compared to plain prompt engineering.
However, recent text-to-image generative models, such as DALLE-3~\cite{betker2023improving}, implicitly use LLMs to suggest diversified prompts to generate the image. 
We compared Varif.ai against two more advanced baselines using GPT-4o~\cite{achiam2023gpt} and Promptist~\cite{hao2024optimizing} as LLMs to prompt the same diffusion model as before, SD-XL Lightning~\cite{lin2024sdxl}.
For each scenario, we prompted the two models to \qquote{provide diverse prompts to generate images of [scenario]} and generated 50 image prompts to gemerate 50 images with the diffusion model. This matches the augmented sample for each participant in the summative study.
We measured the image diversity of each image set to indicate the open-ended diversity, and diversity alignment to indicate how well the image set satisfied the user specifications in terms of attributes.

\textit{Open-Ended Image Diversity.}
Fig.~\ref{fig:promptist} compares computed image diversity of each Varif.ai user participant to the automatic diversity baselines. 
While these automated baselines had higher diversity than most users, some Varif.ai participants did achieve even higher diversity (Fig.~\ref{fig:promptist}a, b), or only marginally lower (Fig.~\ref{fig:promptist}c). 
Although automatic diversity can be effective, it does not necessarily support specific diversity needs of users, which we examine next.

\textit{User-Defined Diversity Alignment.} 
We evaluated how automated diversification baselines compare with Varif.ai in achieving user-defined diversity. 
To standardize the comparison, we focused on the three most common attributes by participants in the summative user study (Fig.~\ref{fig:attributes}), LLM-generated 5 labels for each attribute, and set the label distribution to be uniform.
To measure the actual counts of attribute labels, the first author manually annotated each image, hence the number of attributes analyzed was small for this study.
With these counts, we measure the diversity alignment of each attribute, and average them for each diversification method.

\begin{figure}[t]
  \centering
  \includegraphics[width=6.5cm]{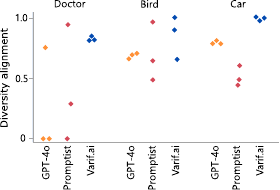}
  \caption{Diversity alignment to uniform attribute distributions of images generated with user-driven Varif.ai (blue), automated diversified
  Promptist~\cite{hao2024optimizing} (red) and GPT-4o~\cite{achiam2023gpt} (orange) for different scenarios. 
  Each point corresponds to one attribute.
  }
  \label{fig:llm_user_aligned}
  \Description{Scatter plot showing distribution alignment scores across three object categories (Doctor, Bird, Car) and three explanation methods (GPT-4o, Promptist, Varif.ai). The y-axis represents distribution alignment (ranging from 0 to 1), and each method-category combination contains multiple colored dots representing individual scores. Varif.ai consistently achieves higher alignment across all categories, with scores clustering near 1.0. Promptist shows more variance and lower alignment in Doctor and Car categories. GPT-4o performs moderately across Bird and Car, but shows poor alignment for Doctor. The figure highlights that Varif.ai explanations more closely match ground-truth distributions across diverse object categories.
  }
\end{figure}

Fig.~\ref{fig:llm_user_aligned} shows that Varif.ai has higher diversity alignment than the automatic baselines, particularly for the Car and Doctor scenarios. 
This indicates that automatic diversity does not target user-driven attribute-based diversity well.
Furthermore, the diversity alignment of automated baselines for the Doctor scenario was especially low (M = 0.37), highlighting the remaining gap to satisfy user-specific requirements of diversity for this socially-sensitive domain.

\subsection{Analysis of CLIP Performance on Diversity}
From our formative study, we noted that the attribute label histograms may be inaccurate.
This could be due to CLIP sometimes mislabeling images. This can affect the performance of the verification and varying steps in Varif.ai.
To investigate the impact of this issue, we conducted a sensitivity analysis on the effect of CLIP labeling accuracy on image diversity measurement and generation, by measuring the diversity alignment based on CLIP labels and ground truth labels, respectively.

\subsubsection{Method.} 
To cover a range of accuracies, we sampled 12 attributes from Fig.~\ref{fig:attributes} to diversify in Varif.ai. 
Attribute labels suggested by the LLM are left unchanged.
Across the scenarios, for each attribute, we generate 20 images with the simple base prompt appended with the attribute and randomly sampled attribute values.
We set Varif.ai to vary the labels to have a diverse uniform distribution in one iteration. 
With the verification step, we note the CLIP-based attribute label (\textit{CLIP label}) that were measured for each image.
The distribution of attribute labels may still not be uniform, indicating a limit in measurement or manipulation which we investigate here.
For each image, we obtain the ground truth attribute label (\textit{Actual label}) with the first author's manually annotation.
Independent variable \textit{CLIP label accuracy} is calculated as how often the CLIP label matches the Actual label.
Dependent variables for \textit{Diversity alignment} based on CLIP or Actual labels are calculated as the KL divergence between a uniform attribute label distribution and the respective distributions.

\subsubsection{Results.}
\begin{figure}[t]
  \centering
  \includegraphics[width=6.5cm]{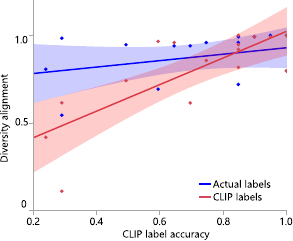}
  \caption{Effect of CLIP-based image-attribute classification accuracy on diversity alignment calculated by CLIP-based labels (red) or Actual labels (blue). 
  Lines show the linear fit with 95\% confidence interval (shaded area).}
  \label{fig:clip_acc}
  \Description{Scatter plot showing the relationship between CLIP label accuracy (x-axis) and diversity alignment (y-axis), with two linear trend lines: one for CLIP-based labels (red) and one for actual labels (blue). Red and blue shaded areas represent 95\% confidence intervals. The red line shows a steeper positive trend, suggesting that higher CLIP accuracy correlates more strongly with increased diversity alignment when using CLIP labels..
  }
\end{figure}

Fig.~\ref{fig:clip_acc} shows how diversity alignment changes with CLIP label accuracy for CLIP-based and Actual labels.
As expected at high accuracies ($>$0.8), both types of alignments are high.
We performed a linear fixed effects regression to determine if the effects of CLIP accuracy are significant.
CLIP-based alignment decreases significantly with decreasing CLIP accuracy ($\beta = 0.76$, $p < .001$), indicating that measurement accuracy of labels decreases as CLIP accuracy is lower.
In contrast, Actual alignment does not significantly change ($\beta = 0.18$, $p = n.s.$), suggesting that the diverse generation is successful regardless of CLIP measurement accuracy.
Perhaps, this is due to the diffusion process using CLIP on the full prompt appended with the attribute label to ensure convergence of the generated image, while our verification step uses only CLIP between the image and the short-text attribute label alone.
Future work could verify based on the long-form appended prompts instead.

\section{Discussion}

We have introduced Varif.ai to help users generate images that are diverse and aligned with their defined attribute distributions.
We discuss its scope, limitations, future work, and generalization.

\subsection{Toward User-Controllable Diversity}
\label{disc:1}
The need for diversity has driven several methods to implicitly diversify generated images~\cite{achiam2023gpt, hao2024optimizing}. However, these methods predetermine the criteria for diversity, which may be too open-ended and exclude specific requirements from users.
Varif.ai supports user-specified diversity based on attributes and targeted label distributions. However, in our formative study, some participants lacked knowledge or goals to diversify effectively when unguided; they did not utilize the feature for \textit{suggesting attributes}, which was shown to be useful in text-based LLMs for proposing criteria to generate perspectives~\cite{hayati2024diversity} and proposing dimensions for creative writing~\cite{suh2023structured}.

\subsection{Scalability to More Attributes and Images}

We explored controlling diversity with a limited number of attributes at a time. Under the open-ended specification degree, summative study participants diversified up to 9 attributes, reflecting both their preferences and time constraints. While participants in the formative study did not express a need to control many attributes, this capability could be crucial for more complex tasks, such as ensuring fairness or creating datasets. Below, we discuss how to scale the processes of verifying and varying diversity across a larger number of attributes.

Verifying diversity across many attributes poses a significant interaction challenge, requiring users to interpret multiple distributions. One potential solution is progressive disclosure: presenting a single summary metric per attribute and revealing detailed histograms only upon user interaction. As in Lyu et al.~\cite{lyu2023if}, diversity metrics can help users quickly assess which attributes have been sufficiently diversified, allowing users to focus on problematic ones. Future work should explore scalable interaction designs that support effective engagement with a large number of distributions.

Our probabilistic prompting approach is scalable to any number of attributes. However, as more attributes are sampled and appended to the prompt, the resulting prompt length can exceed what current image generation models can handle. 
Addressing this may involve improving the capacity of the image generator or refining the prompt strategy, e.g., by sampling a subset of attributes to make the prompt brief while preserving some diversity.

Managing many attributes requires sufficient images to diversify. While our main experiments were limited to 20 images, we conducted preliminary tests with three users generating 100–200 images using Varif.ai. The images were generated within 90 seconds, and participants reported no major difficulties. Histograms and brushing interactions remained effective for verifying diversity even with large image sets. 
For instance, users employed brushing to locate images with underrepresented labels, enabling them to verify concept coverage.
Future work should explore interaction strategies that scale with both the number of attributes and the number of images, enabling users to effectively verify and adjust attribute distributions without sacrificing usability.

\subsection{Improving User Control of Image Sets}
In our user studies, several participants expressed a need for more control over image collections and improved modeling.

\subsubsection{Multimodal Inputs for Image Collection}
Participants wanted greater control over the images, including the ability to modify their own images. While text-based descriptions were useful, they were often insufficient.  Text can be ambiguous, and even precise captions may produce unintended results~\cite{hutchinson2022underspecification}. Concepts such as layout or composition are especially difficult to convey through text alone.
Although technical solutions like layout control exist~\cite{zheng2023layoutdiffusion, hertz2022prompt}, they require new interaction designs to scale effectively for image collections. Future work should explore multimodal input mechanisms to improve user control over image collections.

\subsubsection{Control Multi-Attribute Entanglement}
From our formative study, participants desired more nuanced control over how attributes interact. This parallels the technical challenge of \emph{entanglement} in image generation. 
Disentangling latent spaces remains a key objective in improving generative model interpretability and editability~\cite{wu2023uncovering}, and recent work has begun exposing these concepts to users through interactive interfaces~\cite{evirgen2023ganravel}. 
Our approach treated attributes as independent, assuming that concepts were fully disentangled. 
To provide control over the attributes' relationship, we could explore more sophisticated visualization techniques. 
While histograms assume attributes behave independently, future systems could incorporate richer visualizations---such as high-dimensional correlation matrices~\cite{cutura2020comparing, elmqvist2008rolling}---to reveal dependencies between attributes.
However, this may increase the complexity of interactions and limit accessibility for non-expert users.

Some participants also requested the ability to exclude specific attributes or labels from generation. A possible solution would be to incorporate negative prompts in the user interface~\cite{ban2025understanding},  for example by allowing users to set the probability of a label to zero, effectively blacklisting it from appearing in generated images.

\subsubsection{Associate Attributes to Parts of the Image}
Attributes served as an intuitive mechanism for verifying and varying diversity. However, some participants reported difficulty in linking attributes to parts of the image, such as changing the color of the frogs rather than the background. 
While rephrasing attributes (e.g., using ``frog color'' instead of ``color'') can help, this may not generalize. Future work should explore techniques for explicitly binding attributes to image regions, such as alternative strategies for constructing probabilistic prompts beyond simple concatenation to the context.

\subsubsection{Improve Classification Accuracy for Distribution Verification}
We have exclusively used CLIP~\cite{radford2021learning} for image representation, despite its known biases~\cite{agarwal2021evaluating, wolfe2022american}. Although CLIP provided acceptable accuracy in our tests, misclassifications can undermine trust in the visualizations and skew diversity assessments. 
Future work should consider more robust classification or ensemble methods, and incorporate representations sensitive to emotional tone~\cite{wang2023reprompt} or structural similarity~\cite{lee2023revisiting}. Enhancing classifier reliability is critical for ensuring accurate and trustworthy diversity verification.

\subsection{Generalization to Other Models and Media}
Although we have implemented Varif.ai with one image generation model and text processor, our framework is agnostic to these underlying models. Future work can investigate using Varif.ai as a benchmark platform to compare the diversity of image generation models, and to assess their amenability toward varying diversity.

While our focus has been on image generation, there is a growing need to support diversity in text generation as well, such as for improving the impact of motivational messages~\cite{cox2021directed, wang2022interpretable}, increasing educational value by offering diverse course descriptions~\cite{lee2022promptiverse}, or supporting dataset evaluation~\cite{reif2023visualizing}. Adapting user-defined attribute control to text media may be feasible with interface-level adjustments.
Varif.ai architecture would need minimal changes to adapt to other domains, as it remains agnostic to the underlying large language model, image generator, and classifier. This flexibility enables integration with newer or more specialized models as needed. While we currently host open-source models on a server, Varif.ai could also run on a variety of setups by leveraging API calls, making it adaptable to different computing environments.

\section{Conclusion}

As AI facilitates the creation of images for effective communication, it is important to ensure that the images are generated diversely, for fairness and for creative ideation. 
Varif.ai represents a significant advancement in enabling user-defined control over the diversity of generative images by providing an interactive visualization for defining, verifying, and varying attributes and their labels.
The results from our formative study demonstrate that Varif.ai allows users to diversify image sets in real use cases, and our summative study shows Varif.ai outperforms traditional prompt-engineering in terms of generated diversity and engagement. 
Future work can explore refining the system’s scalability and extending its applicability to more types of data such as text. Varif.ai presents a promising direction for ensuring diverse, user-driven outputs, with potential for further enhancements.

\begin{acks}
This research/project is supported by the National Research Foundation, Singapore under its AI Singapore Programme (AISG Award No: AISG2-PhD-2023-01-040-J), and is part of the programme DesCartes which is supported by the National Research Foundation, Prime Minister’s Office, Singapore under its Campus for Research Excellence and Technological Enterprise (CREATE) programme.
\end{acks}

\bibliographystyle{ACM-Reference-Format}
\bibliography{references}

\appendix

\section{Prompts}
\label{app:llm}

\subsection{Prompts for Suggesting Attribute Labels.}

 \textit{System Prompt:} "You are a useful assistant. You give very brief answers, in very few words, no need to be polite, do not provide explanations."
 
\textit{Instruction prompt}: 
"For the extracted attribute $t_a$ in the context of $t_0$, suggest possible labels for the attribute and provide a precise definition. Consider general knowledge or common scenarios for accuracy. Example: for the attribute `color' in the context of `sky', some possible labels are blue, white, grey, orange, and red. For the attribute `ethnicity' in the context of `person', some possible labels are Caucasian, Black, Asian, Hispanic, and Middle Eastern."

\textit{Answer template:} Here are $n_a$ possible labels of attribute $t_a$ in the context of $t_0$: 
for i in range(n-a):  "i."  + [GENERATE 4 TOKENS]

\subsection{Prompts for Suggesting Attributes.}

 \textit{System Prompt:} "You are a useful assistant. You give very brief answers, in very few words, no need to be polite, do not provide explanations."
 
\textit{Instruction prompt}: 
"For the context of $t_0$, suggest possible attributes to diversify.Consider general knowledge or common scenarios for accuracy."

\textit{Answer template:} Here are 3 possible attributes $t_a$ in the context of $t_0$: 
for i in range(n-a):  "i."  + [GENERATE 4 TOKENS]

\section{Additional Results of Participants}
\label{app:iterations}

\begin{figure*}
    \centering
    \includegraphics[width=\linewidth]{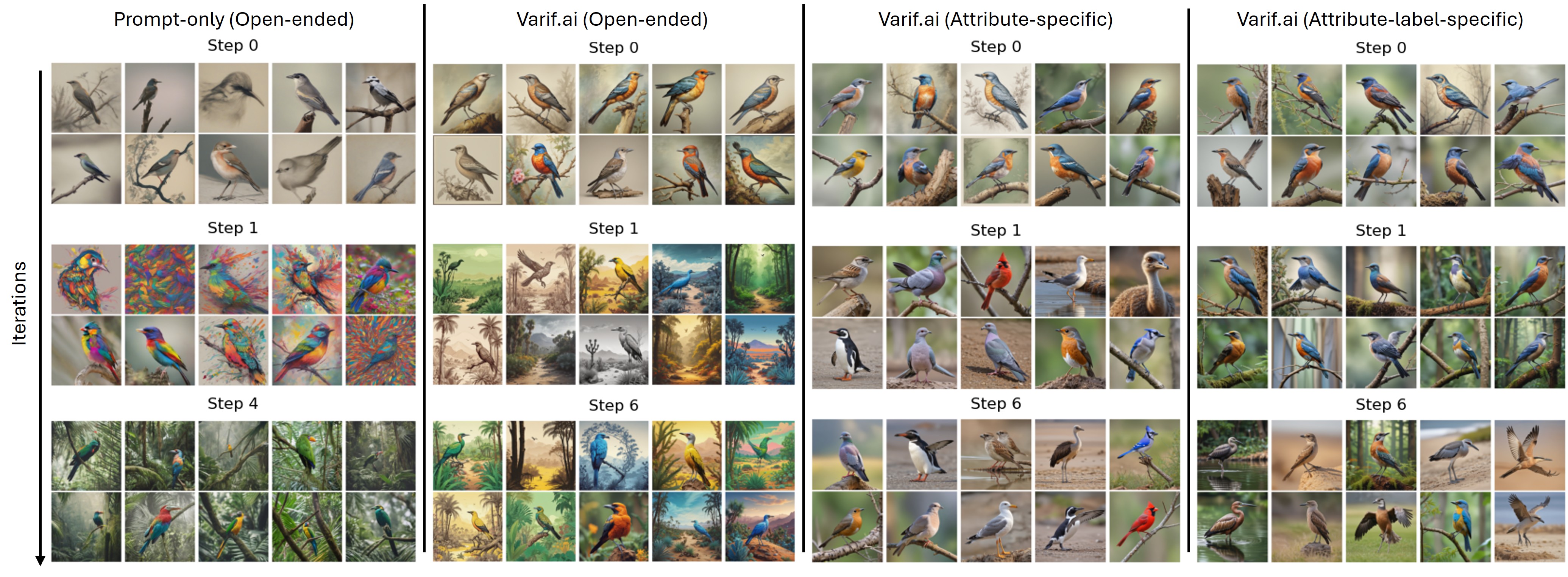}
    \caption{Examples of iterations for different conditions on the \textit{Birds} scenario.}
    \label{app:iter1}
    \Description{"The figure shows examples of iterations for different conditions in the Birds scenario. There are four columns comparing image generation methods: Prompt-only | Open-ended: Displays iterations of generated bird images across several steps, starting from Step 0 to Step 4. The progression shows changes in style and diversity as iterations proceed.Varif.ai | Open-ended: Shows multiple steps (Step 0 through Step 6), where images of birds and landscapes become increasingly diverse and refined as the system iterates. Varif.ai | Attributes: Images progress through steps, with visible changes in bird attributes like color and pose across several iterations, reaching Step 6. Varif.ai | Distributions: The final set of images demonstrates how bird diversity improves over time with adjustments to distributions through steps, showing a more varied set of birds by Step 6. The iterations across these conditions demonstrate the increasing refinement and diversity of the bird images as the system and methods are applied over multiple steps."}
\end{figure*}

\begin{figure*}
    \centering
    \includegraphics[width=\linewidth]{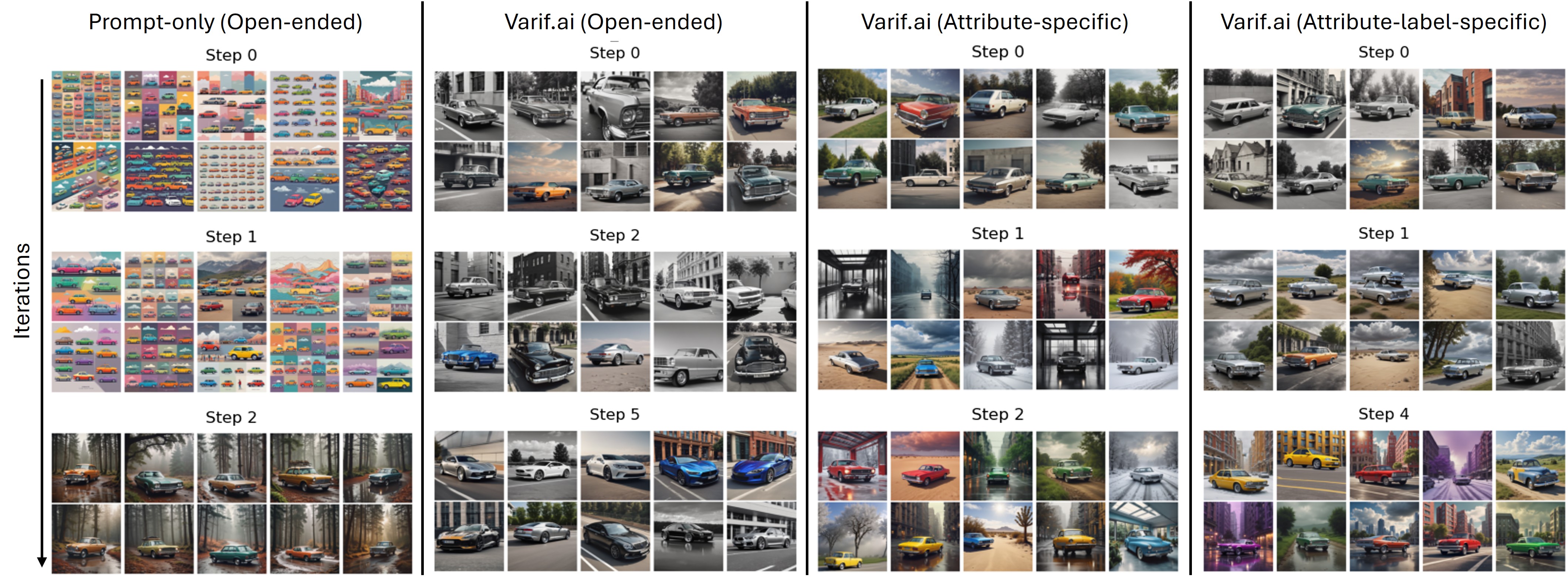}
    \caption{Examples of iterations for different conditions on the \textit{Cars} scenario.}
    \Description{The figure shows examples of iterations for different conditions in the   doctors scenario. There are four columns comparing image generation methods: Prompt-only | Open-ended: Displays iterations of generated   doctor images across several steps, starting from Step 0 to Step 4. The progression shows changes in style and diversity as iterations proceed.Varif.ai | Open-ended: Shows multiple steps (Step 0 through Step 6), where images of   doctors and landscapes become increasingly diverse and refined as the system iterates. Varif.ai | Attributes: Images progress through steps, with visible changes in   doctor attributes like color and pose across several iterations, reaching Step 6. Varif.ai | Distributions: The final set of images demonstrates how   doctor diversity improves over time with adjustments to distributions through steps, showing a more varied set of   doctors by Step 6. The iterations across these conditions demonstrate the increasing refinement and diversity of the   doctor images as the system and methods are applied over multiple steps.}
    \label{app:iter2}

\end{figure*}

\begin{figure*}
    \centering
    \includegraphics[width=\linewidth]{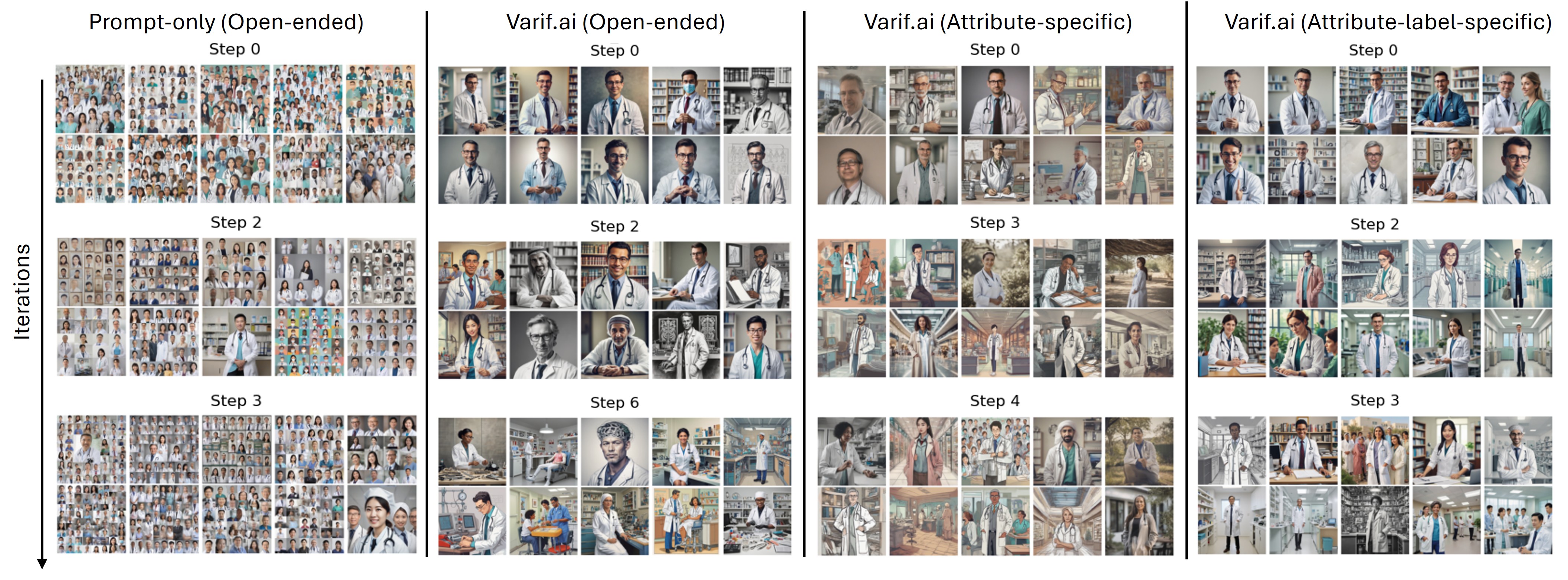}
    \caption{ Examples of iterations for different conditions on the \textit{Doctors} scenario.}
    \label{app:iter3}
    \Description{Grid of four columns showing example image iterations for different generation conditions in the "Doctors" scenario. Each column represents a different method: (1) Prompt-only | Open-ended, (2) Varif.ai | Open-ended, (3) Varif.ai | Attributes, and (4) Varif.ai | Distributions. Each method displays multiple steps (iterations), showing the progression of generated images from initial to refined outputs.
In the first column (Prompt-only), images remain densely populated and visually homogeneous across steps.
The second column (Varif.ai | Open-ended) shows more diverse and realistic portraits of doctors, with changes in ethnicity, gender, and settings over six iterations.
The third column (Varif.ai | Attributes) emphasizes stylistic and contextual variety (e.g., backgrounds, poses) through four steps.
The fourth column (Varif.ai | Distributions) shows balanced and varied representations, including different genders and ethnicities, across three iterations.
Each step illustrates how image content evolves under the influence of different control strategies.}
\end{figure*}

\begin{figure*}
    \centering
    \includegraphics[width=\linewidth]{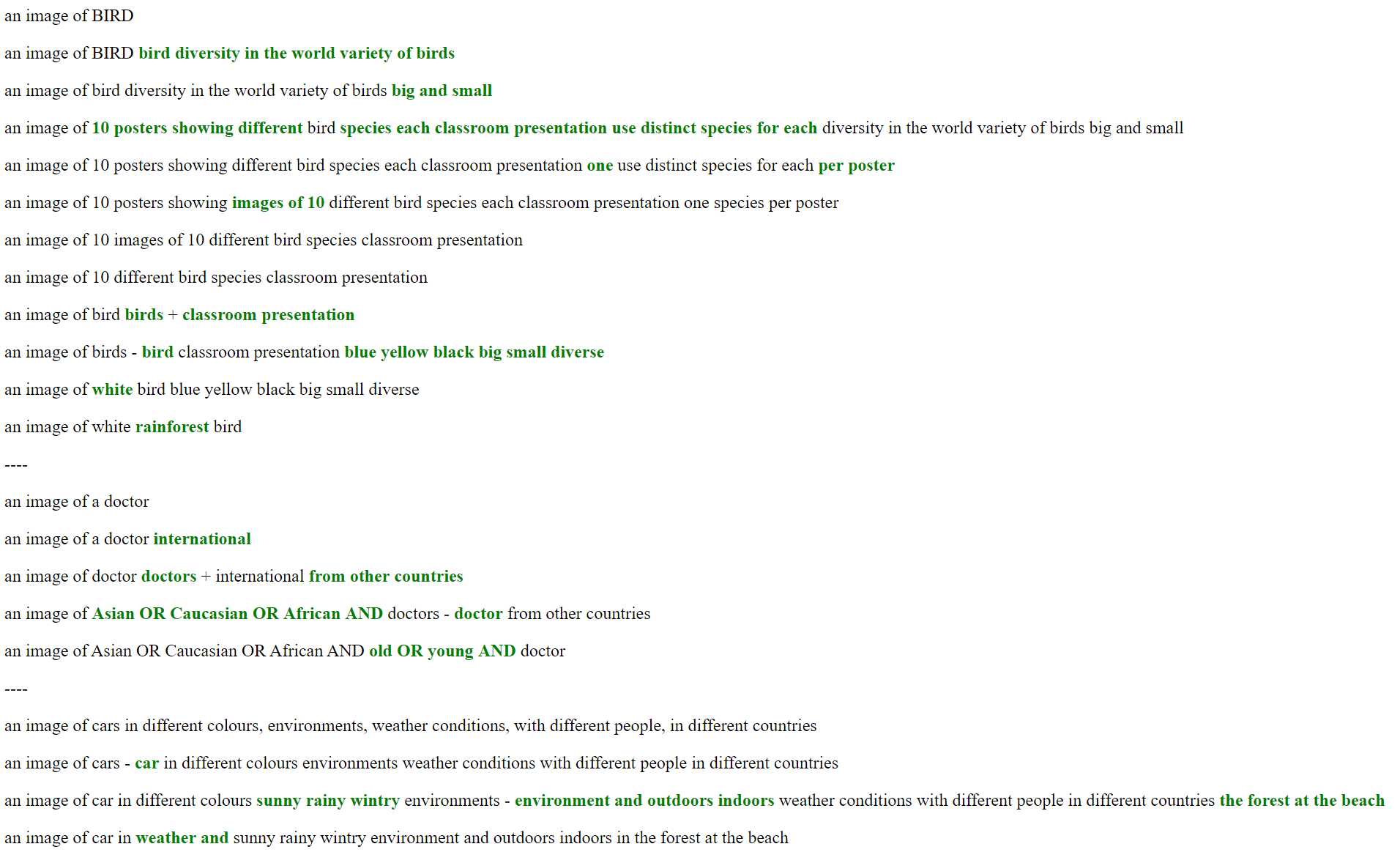}
    \caption{Three examples of prompts iterations for each scenario for the technique Prompt-only. Modifications from one iteration to an other are highlighted in green. }
    \label{app:promptdiff}
    \Description{"The figure displays examples of prompt iterations during three scenarios---Birds, Doctors, and Cars---for the technique Prompt-only.For the Birds scenario, the prompts evolve from general descriptions like "an image of BIRD" to more detailed descriptions incorporating diversity in size, color, species, and classroom presentation.For the Doctors scenario, the prompts start with "an image of a doctor" and gradually become more specific, incorporating international diversity and attributes like age, gender, and ethnicity.For the Cars scenario, the prompts initially focus on "an image of cars" and later expand to include different environments, weather conditions, and people in varied settings.This figure illustrates how participants refined their prompts over multiple iterations to capture more diversity in each scenario."}
\end{figure*}

Fig. ~\ref{app:iter1},~\ref{app:iter2},~\ref{app:iter3} show participants' image collections for the scenarios of the summative study. 
Fig. ~\ref{app:promptdiff} shows participant iterative prompt-engineering during the Prompt-only condition in the summative study.


\end{document}